\documentclass[aps,showpacs,preprintnumbers,amsmath, amssymb]{revtex4}

\usepackage{float}
\usepackage{graphics,epsfig}
\usepackage{graphicx}
\usepackage{dcolumn}
\usepackage{bm}

\begin{document}
\baselineskip=0.8 cm
\title{{\bf Estimation precision of parameter associated with Unruh-like effect}}

\author{Zixu Zhao$^{1}$\footnote{zhao$_{-}$zixu@yeah.net}, Shuhang Zhang$^{1}$\footnote{shuhang$_{-}$zhang@yeah.net}, Qiyuan Pan$^{2}$\footnote{panqiyuan@126.com}, and Jiliang Jing$^{2}$\footnote{jljing@hunnu.edu.cn}}
\affiliation{$^{1}$School of Science, Xi'an University of Posts and Telecommunications, Xi'an 710121, China}
\affiliation{$^{2}$ Key Laboratory of Low Dimensional Quantum
Structures and Quantum Control of Ministry of Education, Hunan
Normal University, Changsha, Hunan 410081, China}

\vspace*{0.2cm}
\begin{abstract}
\baselineskip=0.6 cm
\begin{center}
{\bf Abstract}
\end{center}

We study the quantum Fisher information (QFI) of acceleration, in the open quantum systems, for a two-level atom with the circular motion coupled to a massless scalar field in the Minkowski vacuum without and with a reflecting boundary in the ultra-relativistic limit. As we amplify $a$, the saturation time decreases for $\theta\neq\pi$, but first increases and then decreases for $\theta=\pi$. Without a boundary, there exists a peak value of QFI with a certain time. The QFI varies with the initial state parameter $\theta$, and firstly takes peak value in the ground state of the atom. The variation of QFI with respect to $\theta$ gradually fades away with the evolution of time. With a boundary, the detection range of acceleration has been expanded. The QFI firstly takes the maximum in the excited state of the atom. In addition, we study the QFI of temperature for a static atom immersed in a thermal bath without and with a boundary. The relation between the saturation time and $T$ is similar to $a$. Without a boundary, the QFI of temperature is similar to that of acceleration. With a boundary, the QFI firstly takes peak value in the ground state of the atom, which is different from the behavior of acceleration. The results provide references for the detection of Unruh-like effect.

\end{abstract}

\maketitle
\newpage
\vspace*{0.2cm}

\section{Introduction}

The works of Rindler \cite{Rindler}, Fulling \cite{Fulling}, Hawking \cite{Hawking1,Hawking2}, Davies \cite{Davies1}, DeWitt \cite{DeWitt} and Unruh \cite{Unruh} showed that, in Minkowski spacetime, a no-particle state of inertial observers (the vacuum state) corresponds to a thermal state with temperature $T_{U}=a\hbar/(2\pi c k_{B})$ for uniformly accelerated observers (here, $a$ is the observers' proper acceleration, $\hbar$ is the reduced Planck's constant, c is the speed of light, and $k_{B}$ is the Boltzmann's constant). This is the Unruh effect, which is the result of the quantum field theory \cite{Birrell}. The Unruh effect has been widely studied because of its connection with a number of contemporary research topics, such as cosmological horizons \cite{Gibbons}, thermodynamics of black holes \cite{UnruhWald1,UnruhWald2} and quantum information \cite{Bousso}. Higuchi, Matsas, and Sudarsky analyzed the bremsstrahlung effect associated with a point charge with constant proper acceleration in the frame coaccelerating with the charge \cite{Higuchi,Higuchi2}. Audretsch and M$\ddot{u}$ller studied the contributions of vacuum fluctuations and radiation reaction to the spontaneous excitation of a uniformly accelerated atom in its ground state, which gave an understanding of the role of the different physical processes underlying the Unruh effect \cite{Audretsch}. The decay of accelerated protons reflects the fact that the Unruh effect is mandatory for the consistency of quantum field theory \cite{Matsas,Vanzella}. Along this line, there have been accumulated interest to study the Unruh effect \cite{Benatti,Yu1,Crispino,Hayden,Adami,Jin,Lima,Wang,Tian,Zhao}.

The Unruh effect is usually related with linearly accelerated observers. More realistic situation, the very large acceleration required for experiments is feasible to achieve in the circular motion. It is of great interest to research the circular motion because of the potential significance. For example, in order to generalize the relativistic kinematics of rigid bodies in uniform rectilinear motion to whatever kind of motion, Ehrenfest considered the uniform rotation of rigid bodies and the special theory of relativity \cite{Ehrenfest}. This idea played an important role in the establishment of general relativity.  Letaw and Pfautsch found that the second quantization of the free scalar field was carried out in rotating coordinates and the spectrum of vacuum fluctuations was calculated for an orbiting observer by using these coordinates, which showed that the spectrum of vacuum fluctuations is composed of the usual zero-point energy plus a contribution arising from the observer's acceleration \cite{Letaw}. Bell and Leinaas examined the possibility of using an accelerated spin one-half particle in an external magnetic field as a ``thermometer", to measure the thermal properties of the vacuum fluctuations in an accelerated frame \cite{Bell}, and analyzed further the connection between the Unruh effect and the radiative excitations of electrons in the storage rings \cite{Bell2}. There is a similar effect for the circular uniform acceleration (Similar effect can be called Unruh-like effect), where the excitation spectrum is not purely thermal, but depends on details of the physical system \cite{Letaw,Bell,Bell2}.

Since the direct detection of Unruh-like effect is also difficult, we will study the effect in the frame of the estimation theory which presents the method to obtain the fundamental precision bounds of parameter estimation because of probabilistic and statistical aspects of quantum theory \cite{Helstrom,Holevo}. The estimation error is quantified by the Cram\'{e}r-Rao bound \cite{Helstrom,Cramer} which is inversely proportional to the quantum Fisher information (QFI). A larger quantum Fisher information corresponds to the better precision, which means that finding the larger QFI is very important. Therefore, the QFI has played a significant role in quantum estimation theory and has attracted considerable attention \cite{Buzek,Poli,Li,Giovannetti1,Giovannetti2,Giovannetti3,Sun,Yu,Rajabpour,Gessner,Frowis,Jing}.

The structure of this work is as follows. In Sec. II, we review the QFI and the open quantum system. In Sec. III, for a two-level atom with the circular motion coupled to a massless scalar field in the Minkowski vacuum, we analyze the estimation precision of acceleration by calculating the QFI of acceleration without and with a boundary. In Sec. IV, we study the estimation precision of temperature for a static atom immersed in a thermal bath without and with a boundary. We will summarize our results in the last section.

\section{QFI and open quantum system}
It is well known that, in the quantum metrology, the QFI provides a lower bound to the mean-square error in the estimation by Cram\'{e}r-Rao inequality~\cite{Cramer,Helstrom,Holevo,Braunstein}
\begin{equation}\label{1}
\rm{Var}(X)\geq\frac{1}{N F_X}\;,
\end{equation}
with the number of repeated measurements $N$. Here $F_X$ represents the QFI of parameter $X$, which can be calculated in terms of the symmetric logarithmic derivative operator by
\begin{equation}
F_X=\textrm{Tr}\,(\rho(X)L^2)\;,
\end{equation}
with the symmetric logarithmic derivative Hermitian operator $L$ which satisfies the equation $\partial_X \rho(X)=[\rho(X)L+L\rho(X)]/2$. For a two-level system, the state can be expressed in the Bloch sphere as $\rho=(I+\bm{\omega}\cdot\bm{\sigma})/2$, here $\bm{\omega}=(\omega_1,\omega_2,\omega_3)$ is the Bloch vector and $\bm{\sigma}=(\sigma_1,\sigma_2,\sigma_3)$ denotes the Pauli matrices. Thus, the QFI of parameter $X$ can be written as \cite{Braunstein,Zhong}
\begin{equation}\label{FX}
   F_X=\left\{
    \begin{array}{l}
    \overset{.}|\partial_X\bm{\omega}|^2+\frac{(\bm{\omega}\cdot\partial_X\bm{\omega})^2}{1-|\bm{\omega}|^2}\;,\;\;\,|\bm{\omega}|<1\;,  \\
    |\partial_X\bm{\omega}|^2\;,\;\;\;\;\;\;\;\;\;\;\;\;\;\;\;\;\;\;|\bm{\omega}|=1\;. \\
    \end{array}
    \right.
\end{equation}

The Hamiltonian of a two-level atom system takes the form
\begin{equation}
H=H_s+H_f+H_I\;,
\end{equation}
where $H_s$, $H_f$, and $H_I$ represent the Hamiltonian of the atom, the scalar field, and their interaction. In this work, we only pay attention to the atom and the interaction between the atom and the scalar field. Therefore, we have $H_s=\hbar\omega_0\sigma_3/2$ and $ H_I(\tau)=(a+a_{+})\phi(t,\textbf{x})=\mu(\sigma_{-}+\sigma_{+})\phi(t,\textbf{x})$, where $\omega_0$ is the energy level spacing of the atom, $\phi(t,\textbf{x})$ is the operator of the scalar field, $\mu$ denotes the coupling constant, $\sigma_{+}$ and $\sigma_{-}$ are the atomic raising and lowering operators, respectively.

We express the initial total density matrix of the system as $\rho_{tot}=\rho(0) \otimes |0\rangle\langle0|$, where $\rho(0)$ is the initial reduced density matrix of the atom and $|0\rangle$ is the vacuum state of the field. The evolution of the total density matrix $\rho_{tot}$ obeys
\begin{equation}
\frac{\partial\rho_{tot}(\tau)}{\partial\tau}=-\frac{i}{\hbar}[H,\rho_{tot}(\tau)]\;,
\end{equation}
where $\tau$ is the proper time. Since the interaction between the atom and field is
weak, we rewrite the evolution of the reduced
density matrix $\rho(\tau)$ into the Kossakowski-Lindblad form~\cite{Gorini,Lindblad, Benatti1, Benatti2}
\begin{equation}
\frac{\partial\rho(\tau)}{\partial \tau}= -\frac{i}{\hbar}\big[H_{\rm eff},\,
\rho(\tau)\big]
 + {\cal L}[\rho(\tau)]\ ,
\end{equation}
with
\begin{equation}
{\cal L}[\rho]=\frac{1}{2} \sum_{i,j=1}^3
a_{ij}\big[2\,\sigma_j\rho\,\sigma_i-\sigma_i\sigma_j\, \rho
-\rho\,\sigma_i\sigma_j\big]\ ,
\end{equation}
where the coefficients of Kossakowski matrix $a_{ij}$ are given by
\begin{equation}
a_{ij}=A\delta_{ij}-iB
\epsilon_{ijk}\delta_{k3}-A\delta_{i3}\delta_{j3}\;,
\end{equation}
with
\begin{equation}
A=\frac{\mu^2}{4}[{\cal {G}}(\omega_0)+{\cal{G}}(-\omega_0)]\;,\;~~
B=\frac{\mu^2}{4}[{\cal {G}}(\omega_0)-{\cal{G}}(-\omega_0)]\;.\;~~
\end{equation}
Introducing the two-point correlation function of the scalar field
\begin{equation}\label{G}
G^{+}(x,x')=\langle0|\phi(t,\textbf{x})\phi(t',\textbf{x}')|0 \rangle\;,
\end{equation}
we can define the Fourier and Hilbert transformations of the field correlation function, ${\cal G}(\lambda)$ and ${\cal K}(\lambda)$ as follows:
\begin{equation}
{\cal G}(\lambda)=\int_{-\infty}^{\infty} d\Delta\tau \,
e^{i{\lambda}\Delta\tau}\, G^{+}\big(\Delta\tau\big)\; ,
\quad\quad
{\cal K}(\lambda)=\frac{P}{\pi
i}\int_{-\infty}^{\infty} d\omega\ \frac{ {\cal G}(\omega)
}{\omega-\lambda} \;.
\end{equation}
By absorbing the Lamb shift term, the effective Hamiltonian $H_{\rm eff}$ can be expressed as
\begin{equation}
H_{\rm eff}=\frac{1}{2}\hbar\Omega\sigma_3=\frac{\hbar}{2}\{\omega_0+\frac{i}{2}[{\cal
K}(-\omega_0)-{\cal K}(\omega_0)]\}\,\sigma_3\;.
\end{equation}
Making use of the ansatz the initial state of the atom $|\psi(0)\rangle=\cos\frac{\theta}{2}|+\rangle+e^{i\phi}\sin\frac{\theta}{2}|-\rangle$, we obtain the evolution of Bloch vector
\begin{align}\label{omega}
&\omega_1(\tau)=\sin\theta \cos(\Omega\tau+\phi)e^{-2 A\tau}\;,\nonumber\\
&\omega_2(\tau)=\sin\theta \sin(\Omega\tau+\phi)e^{-2 A\tau}\;,\nonumber\\
&\omega_3(\tau)=\cos\theta e^{-4 A\tau}-\frac{B}{A}\left(1-e^{-4 A\tau}\right)\;.
\end{align}

\section{Quantum estimation of acceleration without and with a boundary}

\subsection{Estimation of acceleration without boundary}

To study the estimation precision of acceleration, we calculate the QFI of acceleration of a two-level atom with the circular motion without boundary. We adopt natural units $c=\hbar=k_{B}=1$. For a two-level atom with the circular motion, the trajectory of the atom can be expressed as
\begin{eqnarray}
t(\tau)=\gamma\tau\ ,\ \ \ x(\tau)=R\cos \frac{\gamma\tau v}{R}\ ,\ \
\ y(\tau)=R\sin \frac{\gamma\tau v}{R}\ ,\ \ \ z(\tau)=0\label{traj}\;,
\end{eqnarray}
where $v$ denotes the tangential velocity of the atom, $R$ is the radius of the orbit, and $\gamma={1}/{\sqrt{1-{v^2}}}$ is the Lorentz factor. The centripetal acceleration of the atom is $a={\gamma^2v^2}/{R}$.

The Wightman function of a massless scalar field in the Minkowski vacuum takes the form

\begin{eqnarray}\label{wightman}
G^+(x,x')_0=-\frac{1}{4\pi^2}
           \frac{1}{(t-t'-i\varepsilon)^2-(x-x')^2-(y-y')^2-(z-z')^2}\;.
\end{eqnarray}
Applying the trajectory of the atom (\ref{traj}), we need expand $\sin^2[a\Delta\tau/(2v{\gamma})]=\frac{a^2{(\Delta\tau)}^2}{4v^2{\gamma}^2}
-\frac{a^4{(\Delta\tau)}^4}{48v^4{\gamma}^4}+\frac{a^6{(\Delta\tau)}^6}{1440v^6{\gamma}^6}-...$ with $\Delta\tau=\tau-\tau'$. Since it is hard to find the explicit form of ${\cal G}(\omega_0)$ and ${\cal G}(-\omega_0)$ with all orders of $\Delta\tau$, we consider the ultra-relativistic limit $\gamma\gg1$~\cite{Bell}, in which
\begin{eqnarray}
G^+(x,x')_0=-\frac{1}{4\pi^2} \frac{1}{(\Delta\tau-i\varepsilon)^2[1+a^{2}(\Delta\tau-i\varepsilon)^2/12]}\;.
\end{eqnarray}

Therefore, the Fourier transformation of the field correlation function is
\begin{eqnarray}\label{fourier}
{\cal G}^0(\omega_0)=\frac{\omega_0}{2\pi}\bigg[1+\frac{a }{4\sqrt{3}\omega_0}e^{-{\frac{2\sqrt{3}\omega_0}{a}}}\bigg]\;.
\end{eqnarray}
Similarly, we have
\begin{eqnarray}\label{fourier2}
{\cal G}^0(-\omega_0)=\frac{\omega_0}{2\pi}\frac{a }{4\sqrt{3}\omega_0}e^{-{\frac{2\sqrt{3}\omega_0}{a}}}\;.
\end{eqnarray}

We can obtain
\begin{eqnarray}\label{A0B0}
\begin{aligned}
&A_{0}=\frac{\mu^{2}\omega_0}{8\pi}\bigg[1+\frac{a }{2\sqrt{3}\omega_0}e^{-{\frac{2\sqrt{3}\omega_0}{a}}}\bigg]\;,\\
&B_{0}=\frac{\mu^{2}\omega_0}{8\pi}.
\end{aligned}
\end{eqnarray}

In the following discussion, we use $\tau\rightarrow \tilde{\tau}\equiv{\mu^2\omega_0}\tau/{(2\pi)}$, $a\rightarrow \tilde{a}\equiv{a}/{\omega_0}$.  For simplicity, $\tilde{\tau}$ and $\tilde{a}$ will be written as $\tau$ and $a$. From (\ref{FX}), (\ref{omega}), and (\ref{A0B0}), we can calculate the QFI of the acceleration $F_a$. It should be noted that $\Omega$ and $\phi$ will disappear in the calculation for $F_{a}$ since $\cos^{2}(\Omega\tau+\phi)+\sin^{2}(\Omega\tau+\phi)=1$.

\begin{figure}[H]
\begin{centering}
\includegraphics[scale=0.55]{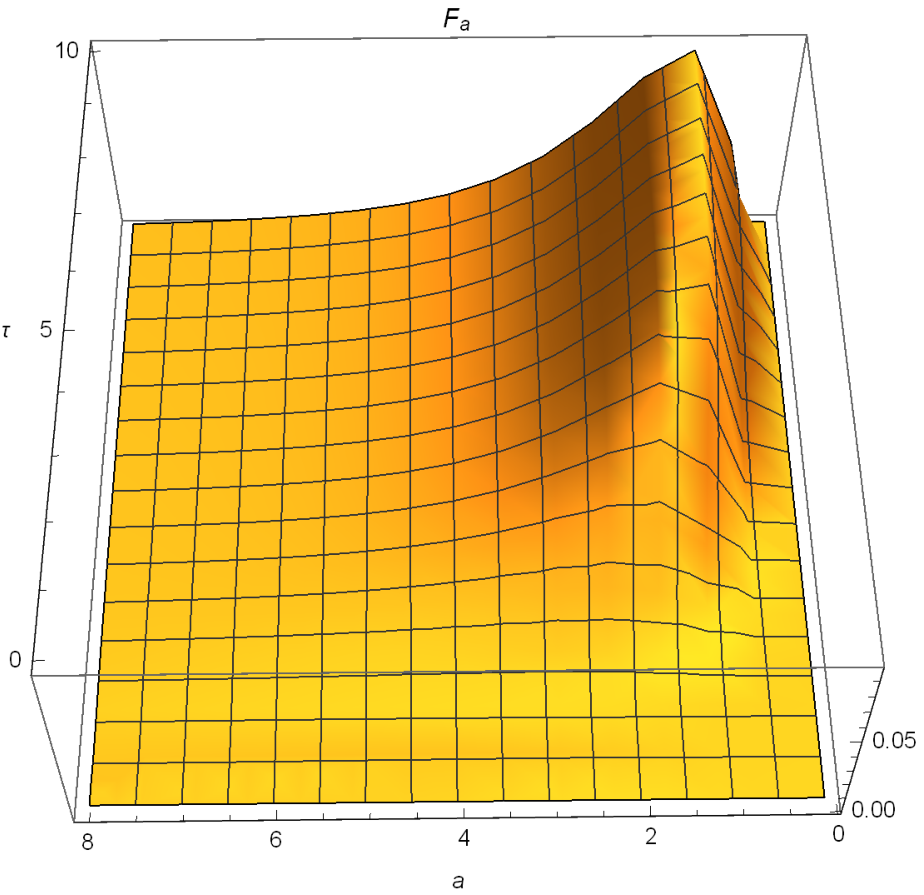}
\includegraphics[scale=0.55]{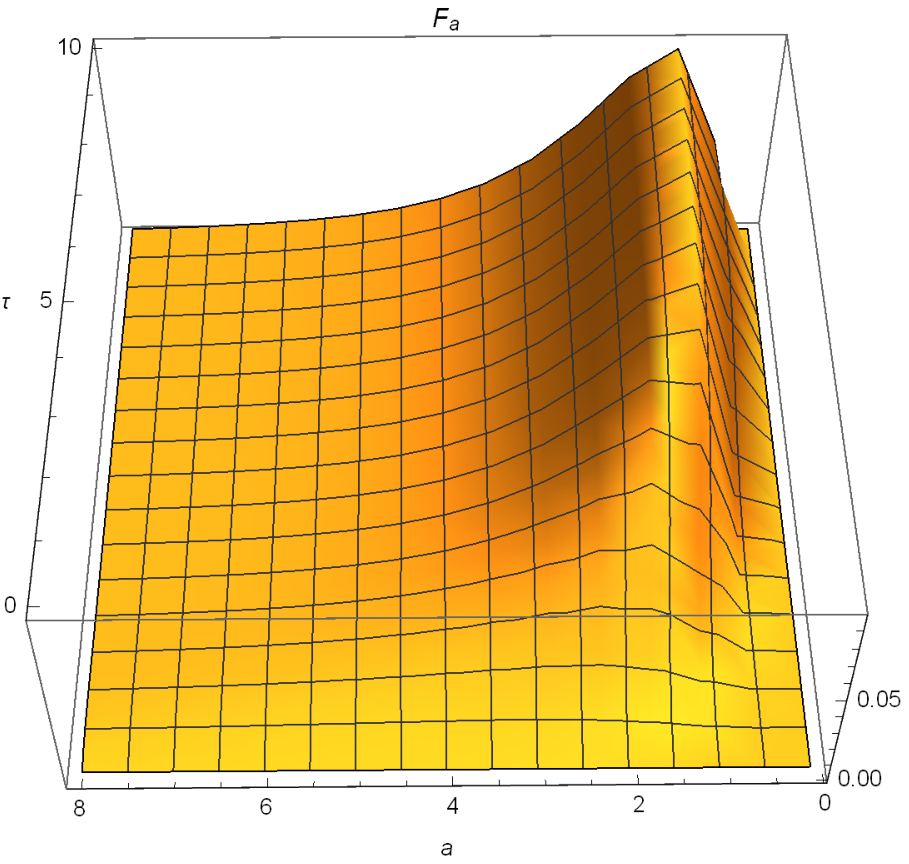}
\includegraphics[scale=0.55]{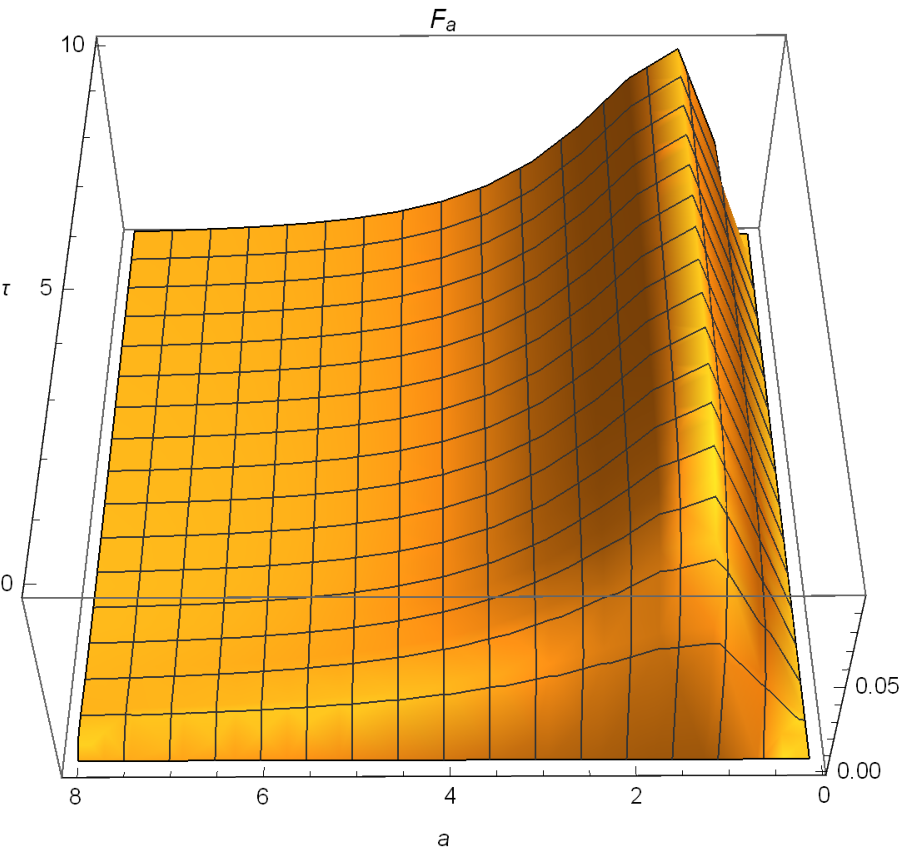}
\caption{\label{fig-Fat} The QFI of acceleration as a function of the acceleration $a$ and the evolution time $\tau$ for different initial state parameters $\theta$. We take $\theta=0$ (left panel), $\theta=\pi/2$ (middle panel), and $\theta=\pi$ (right panel). }
\end{centering}
\end{figure}

\begin{figure}[H]
\begin{centering}
\includegraphics[scale=0.53]{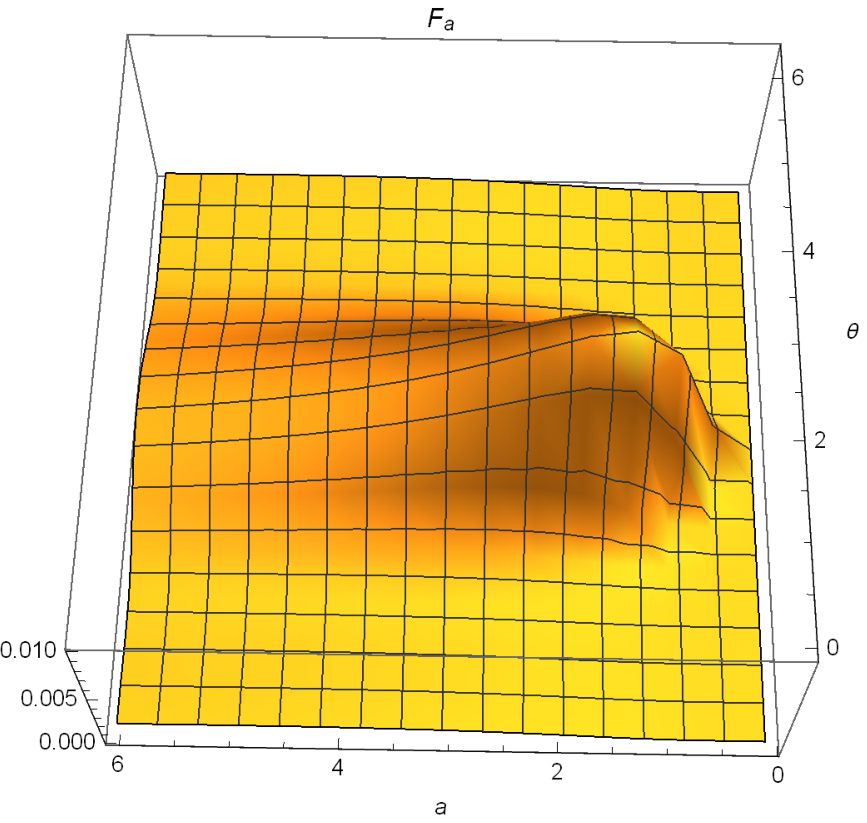}
\includegraphics[scale=0.53]{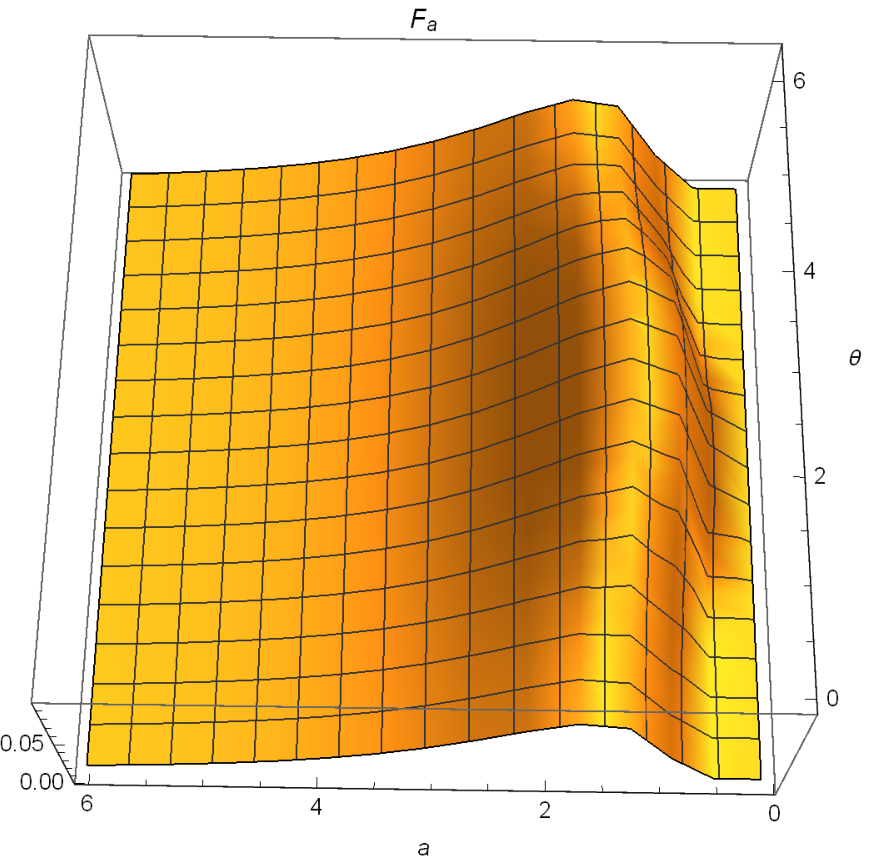}
\includegraphics[scale=0.53]{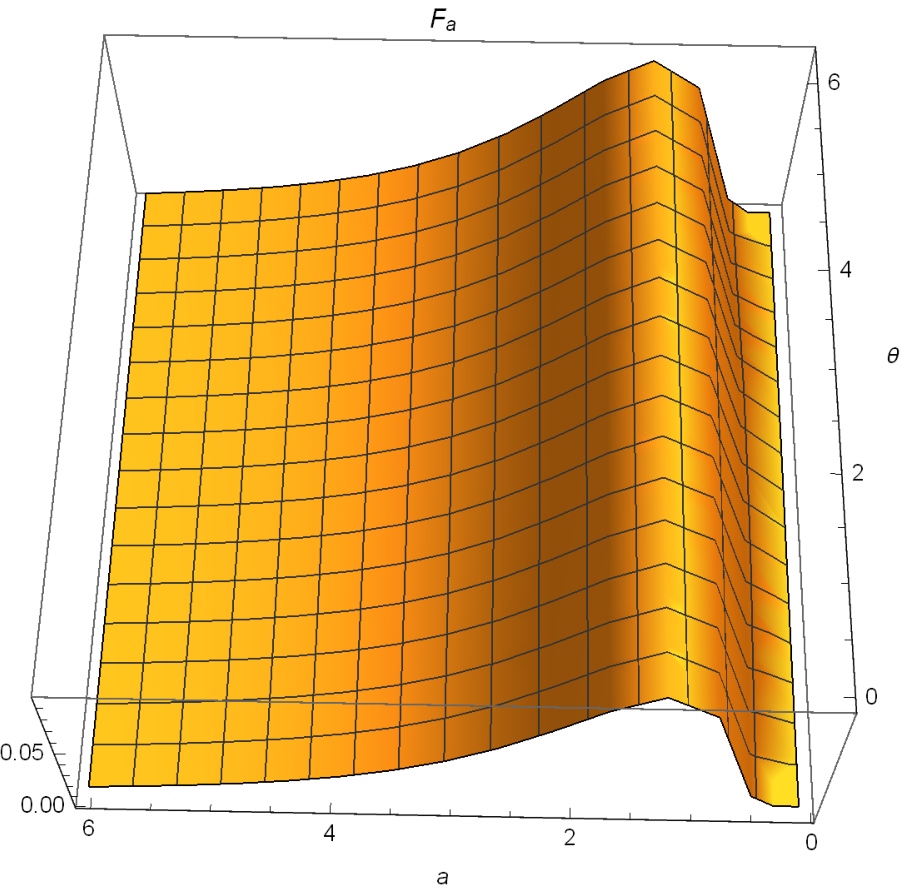}
\caption{\label{fig-Fath} The QFI of acceleration as a function of the acceleration $a$ and the initial state parameter $\theta$ for different evolution time $\tau$. We take $\tau=0.1$ (left panel), $\tau=5$ (middle panel), and $\tau=9$ (right panel). }
\end{centering}
\end{figure}

\begin{figure}[H]
\begin{centering}
\includegraphics[scale=0.75]{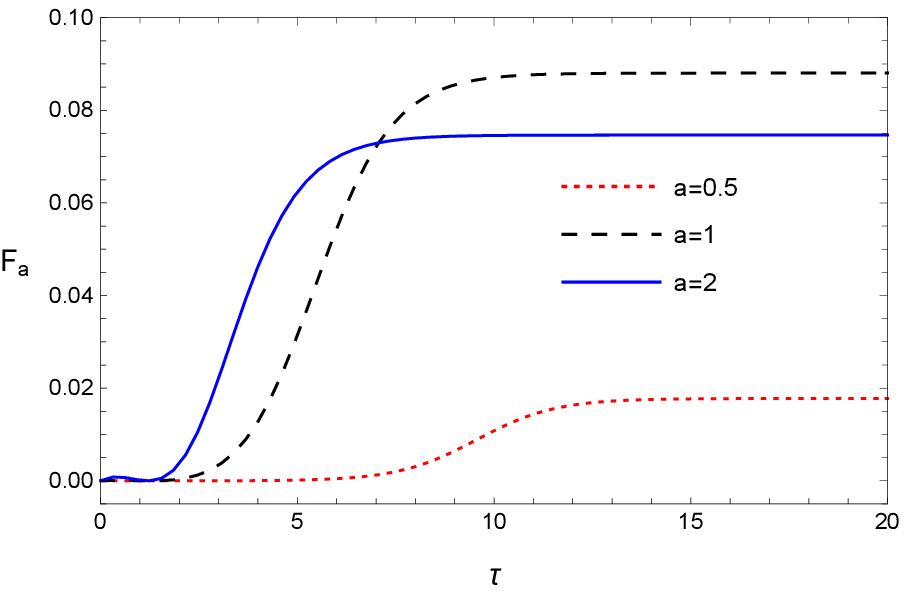}
\includegraphics[scale=0.72]{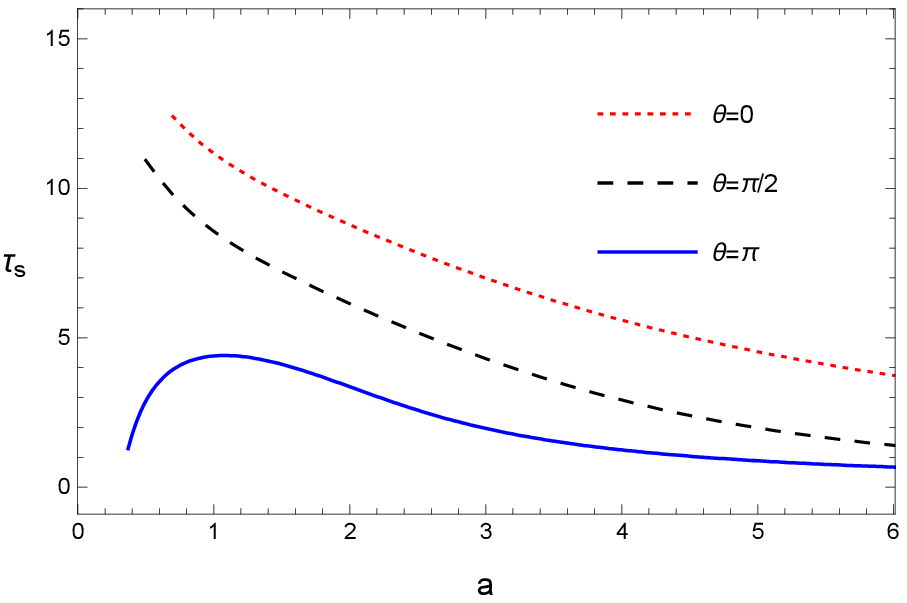}
\includegraphics[scale=0.75]{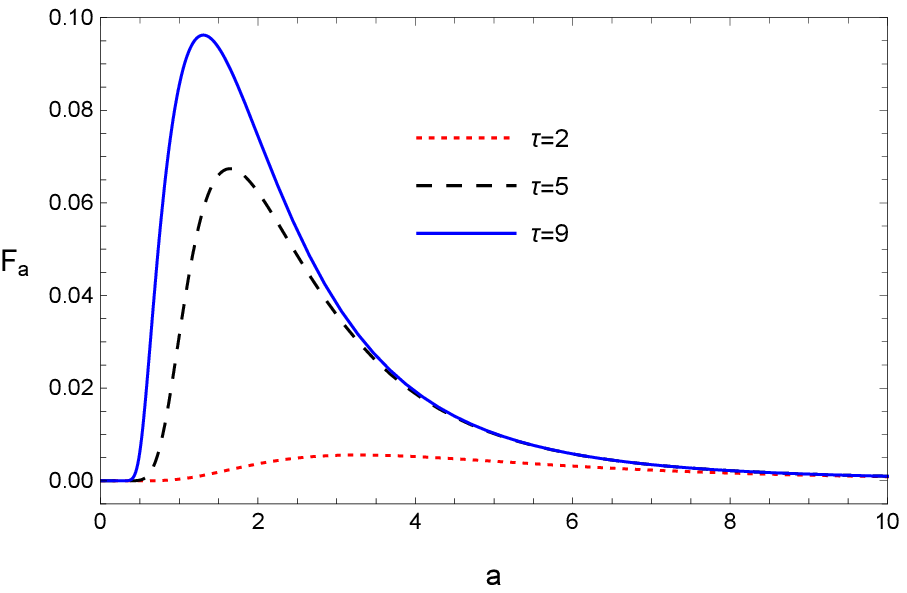}
\includegraphics[scale=0.75]{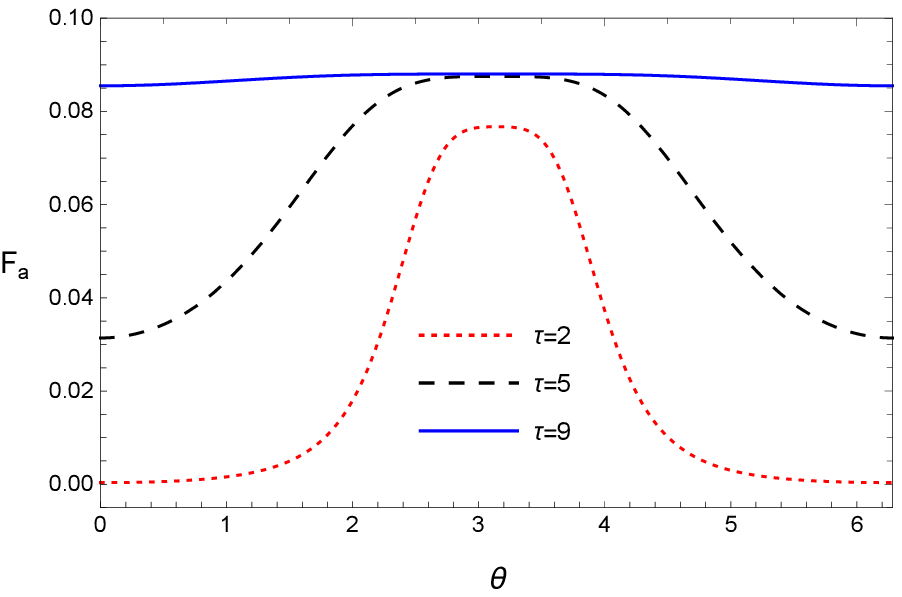}
\caption{\label{2dfig-a}The QFI of acceleration as a function of $\tau$ for $\theta=0$ with different $a$ in the top left panel, as a function of $a$ for $\theta=0$ with different $\tau$ in the bottom left panel, and as a function of $\theta$ for $a=1$ with different $\tau$ in the bottom right panel. The saturation time as a function of $a$ with different $\theta$ by using numerical method in the top right panel.}
\end{centering}
\end{figure}

In an unbounded space, we obtain the QFI of acceleration $F_a$ as a function of $a$ and $\tau$ for $\theta=0$, $\theta=\pi/2$, and $\theta=\pi$ in Fig. \ref{fig-Fat}. For the fixed acceleration, we observe that $F_a$ can reach stable value when the atom evolves for some time. $F_a$ first increases and then decreases as we amplify the value of acceleration with a certain time. Furthermore, we plot the QFI with different $a$ as a function of $\tau$ for fixed $\theta=0$ in the top left panel of Fig. \ref{2dfig-a}. We find that the QFI with $a=2$ reaches stable value which is also the maximum value faster than $a=1$ and $a=0.5$ for $\theta=0$. We define the saturation time as the minimum time when the QFI reaches the stable value. With increase of $a$, the saturation time decreases for $\theta\neq\pi$, but first increases and then decreases for $\theta=\pi$. We just present the case of $\theta=0$, $\theta=\pi/2$, and $\theta=\pi$ in the top right panel. In the bottom left panel of Fig. \ref{2dfig-a}, the peak value of QFI increases with the evolution of time, and eventually reaches to the maximum. For fixed $\tau$, while $a$ is larger than the specific value which corresponds to the peak value of QFI, the precision reduces. Considering the complex expression of QFI, we give these specific values by the numerical method. We obtain the specific values $a=3.2822$ for $\tau=2$, $a=1.6448$ for $\tau=5$, and $a=1.3055$ for $\tau=9$. In Fig. \ref{fig-Fath}, we plot the QFI of acceleration $F_a$ as a function of $a$ and $\theta$ for $\tau=0.1$, $\tau=5$, and $\tau=9$. From the left panel, we find that the QFI varies with the initial state parameter $\theta$, and $F_a$ takes peak value at $\theta=\pi$ which corresponds to the ground state of the atom. However, the variation of QFI with respect to $\theta$ gradually fades away with the evolution of time as shown in the middle panel and right panel. In the bottom right panel of Fig. \ref{2dfig-a}, $F_a$ goes to the maximum for any $\theta$ beyond a certain time. Therefore, the optimal precision of estimation is achieved when choosing an appropriate range.

\subsection{Estimation of acceleration with a boundary}

We add a boundary at $z=0$ and consider an atom moving in the $x-y$ plane at a distance $z$ from the boundary. Then, the two-point function in this case is expressed as
\begin{eqnarray}
G^+(x,x')
=G^+(x,x')_0+G^+(x,x')_b\;,
\end{eqnarray}
where $G^+(x,x')_0$ is the two-point function in the unbounded case which has already been calculated above, and
\begin{eqnarray}
G^+(x,x')_b&=&-\frac{1}{4\pi^2}
\frac{1}{(x-x')^2+(y-y')^2+(z+z')^2-(t-t'-i\varepsilon)^2},\;
\end{eqnarray}
gives the correction due to the presence of the boundary. Similarly, we have
\begin{eqnarray}
G^+(x,x')=-\frac{1}{4\pi^2} \frac{1}{4z^2-(\Delta\tau-i\varepsilon)^2-a^2(\Delta\tau-i\varepsilon)^4/12}\;.
\end{eqnarray}
The Fourier transformations of the correlation function can be written as
\begin{eqnarray}\label{zheng}
{\cal G}(\omega_0)&=&\frac{\omega_0}{2\pi}\bigg[1+\frac{a }{4\sqrt{3}\omega_0}e^{-{\frac{2\sqrt{3}\omega_0}{a}}}-\frac{\sqrt{3}a}{\sqrt{-3+\sqrt{9+12a^2z^2}}{\sqrt{6+8a^2z^2}\omega_0}}\sin\frac{\sqrt{-6+2\sqrt{9+12a^2z^2}}\omega_0}{a}\nonumber\\
&-&\frac{\sqrt{3}a}{2\sqrt
{3+\sqrt{9+12a^2z^2}}{\sqrt
{6+8a^2z^2}\omega_0}}e^{-\frac{\sqrt
{6+2\sqrt{9+12a^2z^2}}\omega_0}{a}}\bigg]\;,
\end{eqnarray}

\begin{equation}\label{fu}
{\cal G}(-\omega_0)=\frac{\omega_0}{2\pi}\bigg[\frac{a }{4\sqrt{3}\omega_0}e^{-{\frac{2\sqrt{3}\omega_0}{a}}}-\frac{\sqrt{3}a}{2\sqrt
{3+\sqrt{9+12a^2z^2}}{\sqrt
{6+8a^2z^2}\omega_0}}e^{-\frac{\sqrt
{6+2\sqrt{9+12a^2z^2}}\omega_0}{a}}\bigg]\;.
\end{equation}

We have
\begin{eqnarray}
A_{b}&=&\frac{\mu^{2}\omega_0}{8\pi}\bigg[1+\frac{a }{2\sqrt{3}\omega_0}e^{-{\frac{2\sqrt{3}\omega_0}{a}}}-\frac{\sqrt{3}a}{\sqrt{-3+\sqrt{9+12a^2z^2}}{\sqrt{6+8a^2z^2}\omega_0}}\sin\frac{\sqrt{-6+2\sqrt{9+12a^2z^2}}\omega_0}{a}\nonumber\\
&-&\frac{\sqrt{3}a}{\sqrt
{3+\sqrt{9+12a^2z^2}}{\sqrt
{6+8a^2z^2}\omega_0}}e^{-\frac{\sqrt
{6+2\sqrt{9+12a^2z^2}}\omega_0}{a}}\bigg]\;,\nonumber\\
B_{b}&=&\frac{\mu^{2}\omega_0}{8\pi}\bigg[1-\frac{\sqrt{3}a}{\sqrt{-3+\sqrt{9+12a^2z^2}}{\sqrt{6+8a^2z^2}\omega_0}}\sin\frac{\sqrt{-6+2\sqrt{9+12a^2z^2}}\omega_0}{a}\bigg].
\end{eqnarray}

In the following discussion, we use $\tau\rightarrow \tilde{\tau}\equiv{\mu^2\omega_0}\tau/{(2\pi)}$, $a\rightarrow \tilde{a}\equiv{a}/{\omega_0}$, and $z\rightarrow \tilde{z}\equiv z \omega_0$. For simplicity, $\tilde{\tau}$, $\tilde{a}$ and $\tilde{z}$ will be  written as $\tau$, $a$ and $z$. We can obtain the QFI of the acceleration $F_a$.

\begin{figure}[H]
\begin{centering}
\includegraphics[scale=0.55]{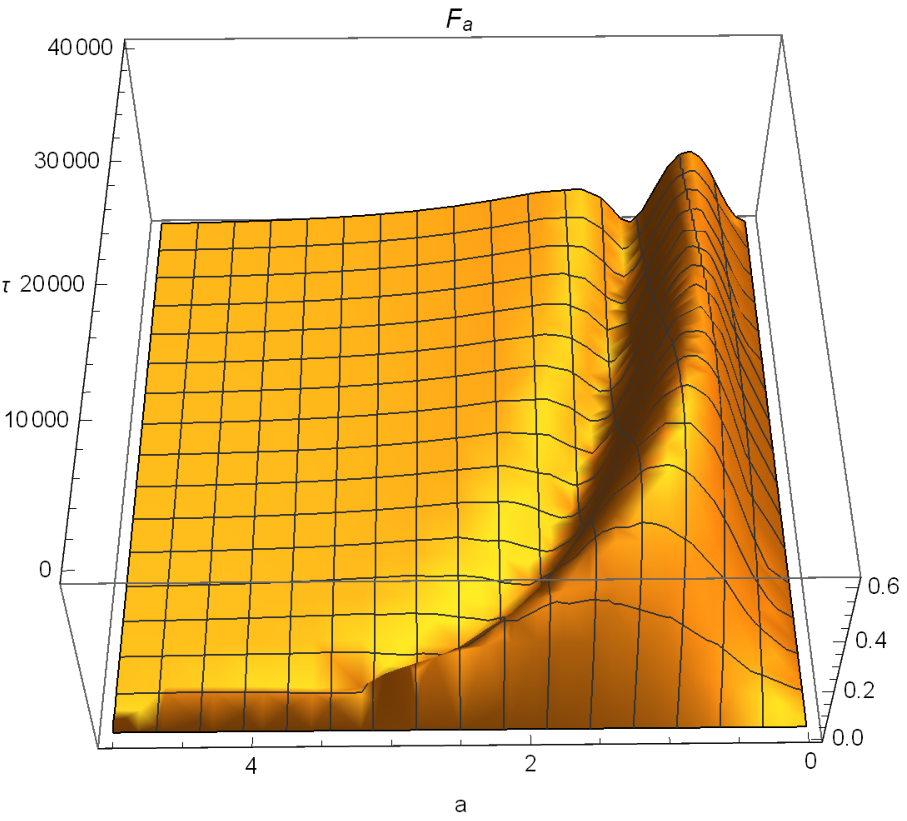}
\includegraphics[scale=0.55]{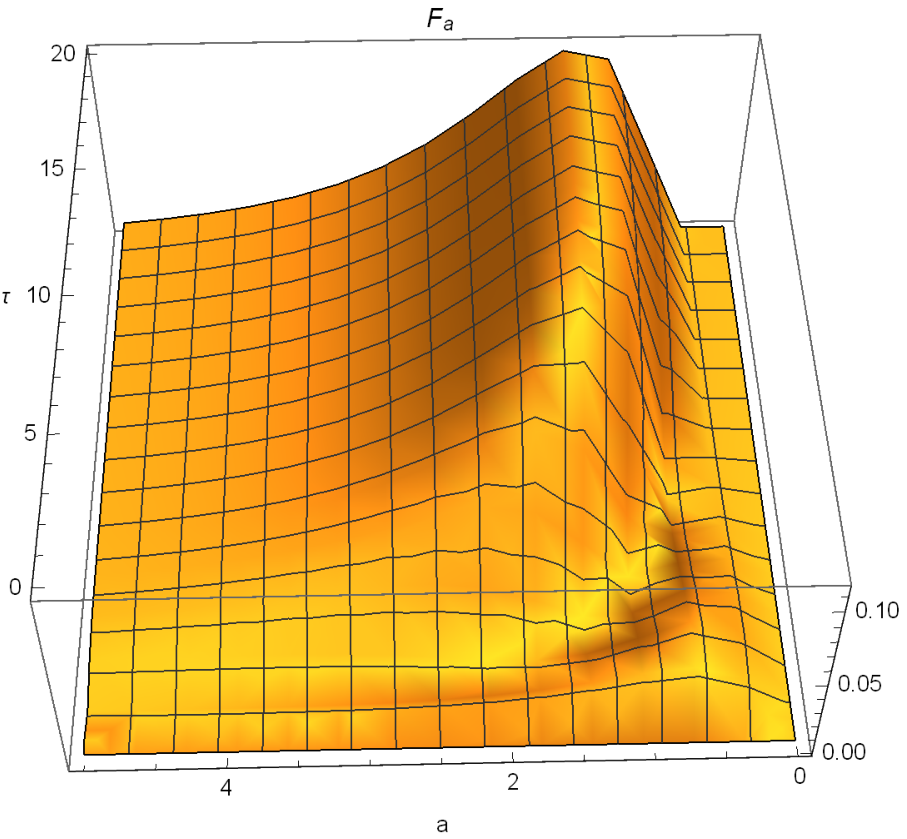}
\includegraphics[scale=0.55]{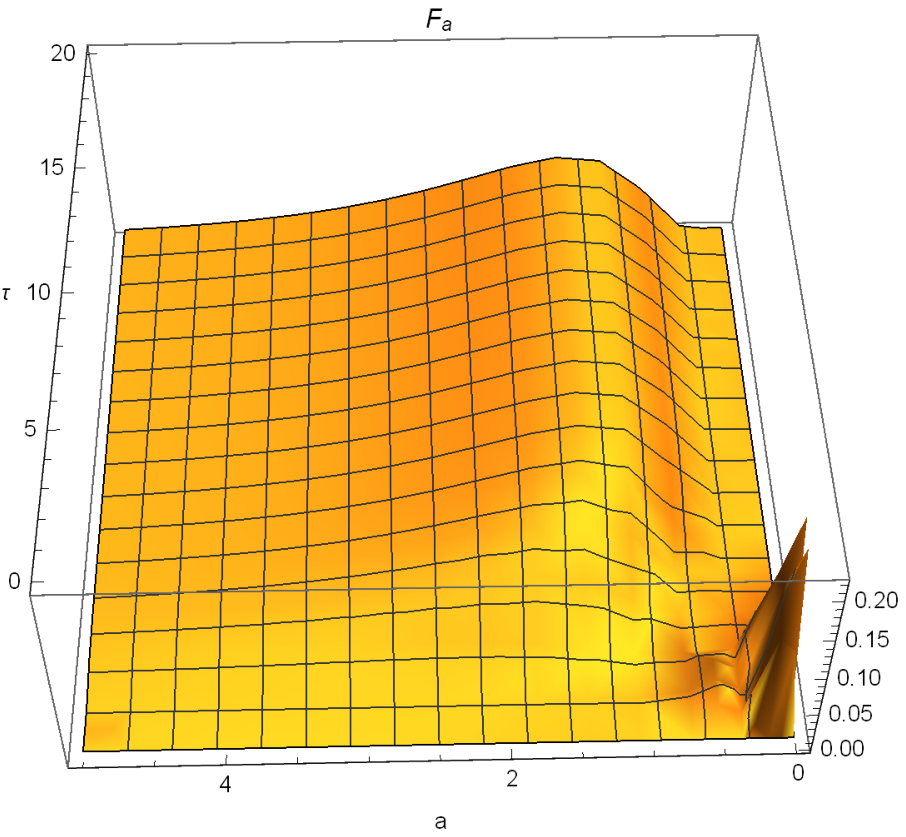}
\caption{\label{fig-Fat-b} The QFI of acceleration as a function of $a$ and $\tau$ for $\theta=0$. We take $z=0.01$ (left panel), $z=1$ (middle panel), and $z=5$ (right panel). }
\end{centering}
\end{figure}

\begin{figure}[H]
\begin{centering}
\includegraphics[scale=0.53]{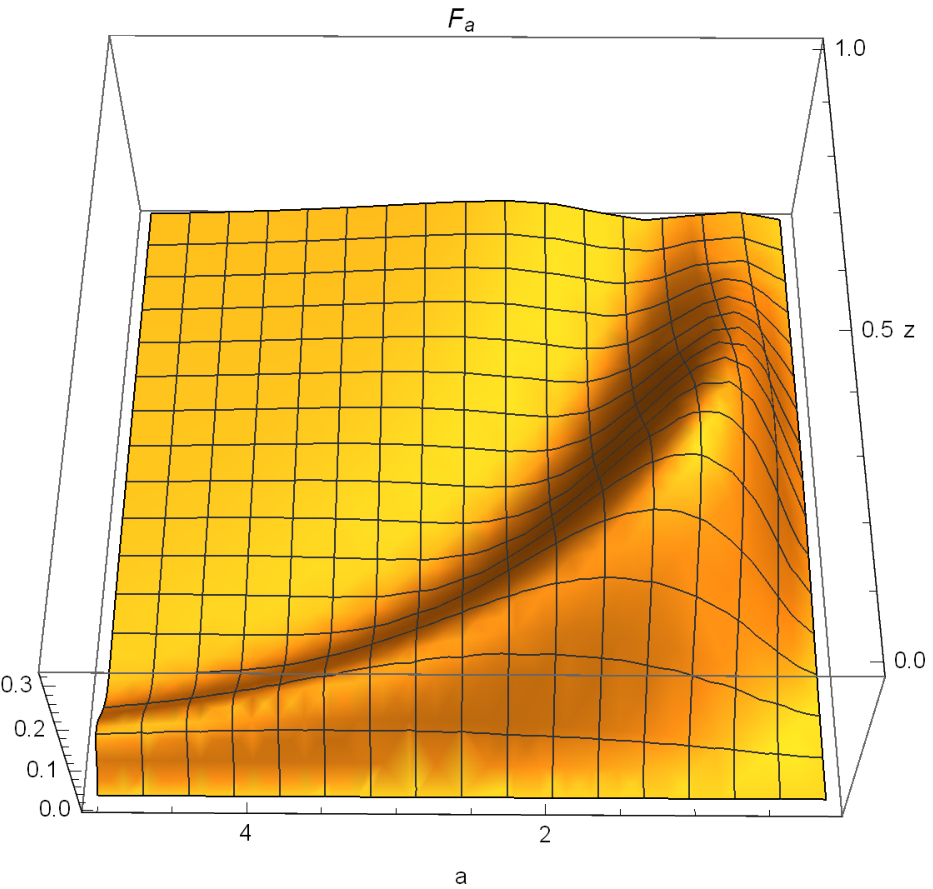}
\includegraphics[scale=0.53]{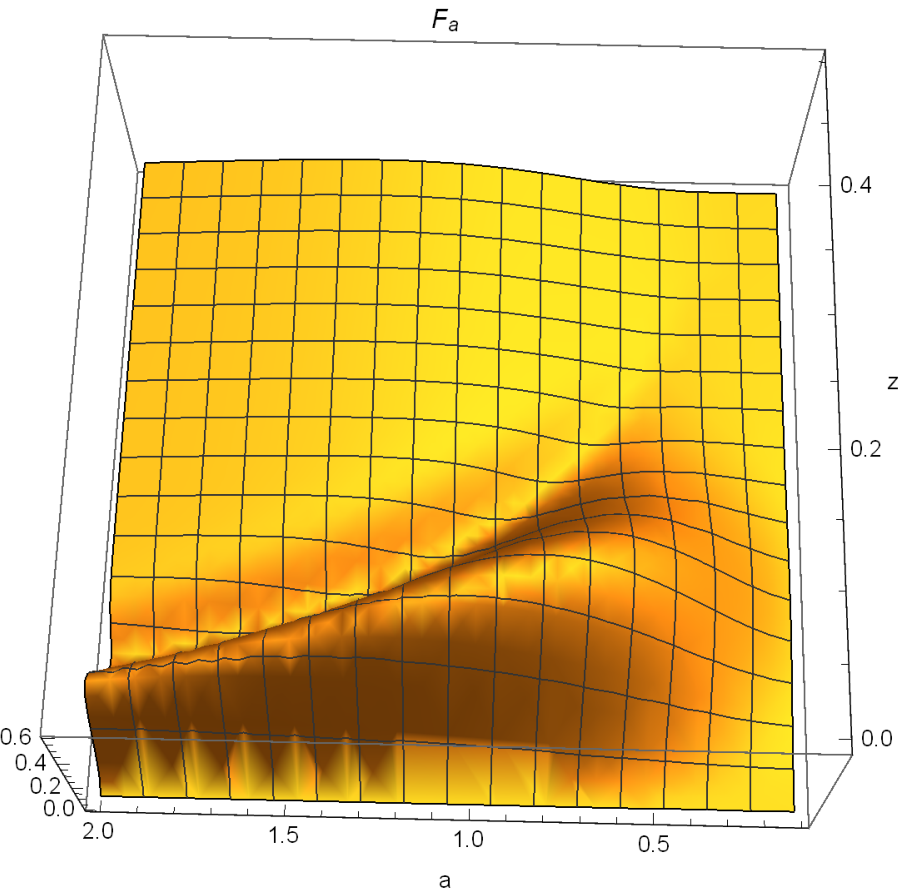}
\includegraphics[scale=0.53]{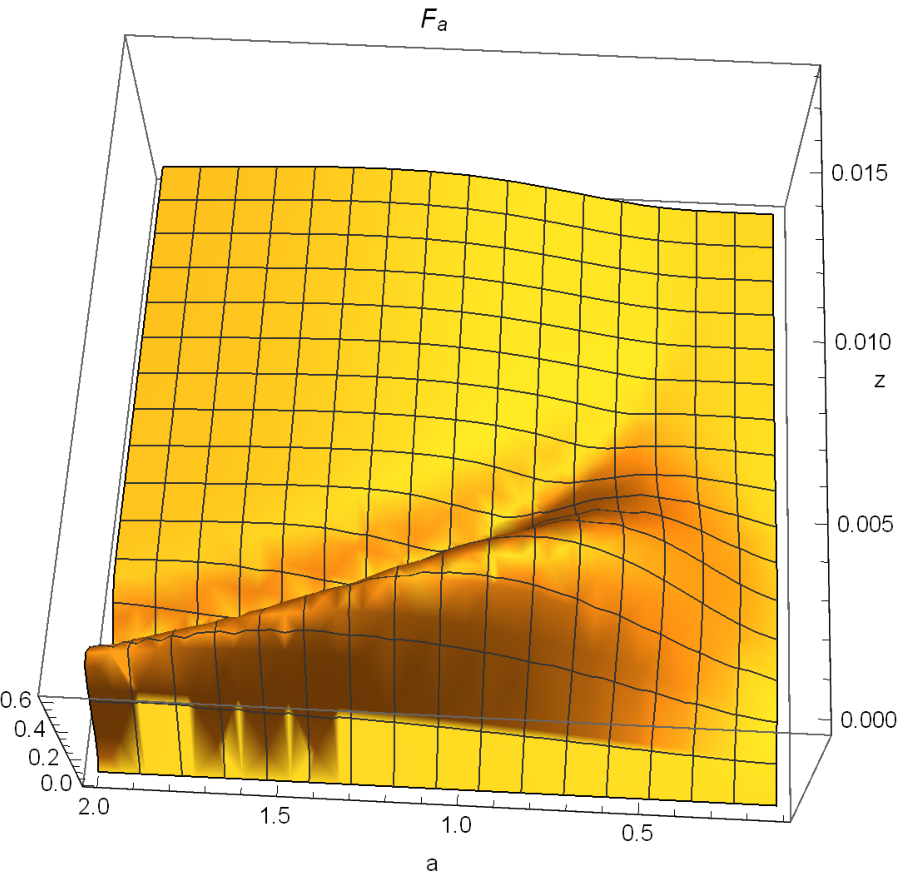}
\caption{\label{fig-Faz-b} The QFI of acceleration as a function of $a$ and $z$ for $\theta=0$. We take $\tau=5$ (left panel), $\tau=100$ (middle panel), and $\tau=80000$ (right panel). }
\end{centering}
\end{figure}

\begin{figure}[H]
\begin{centering}
\includegraphics[scale=0.55]{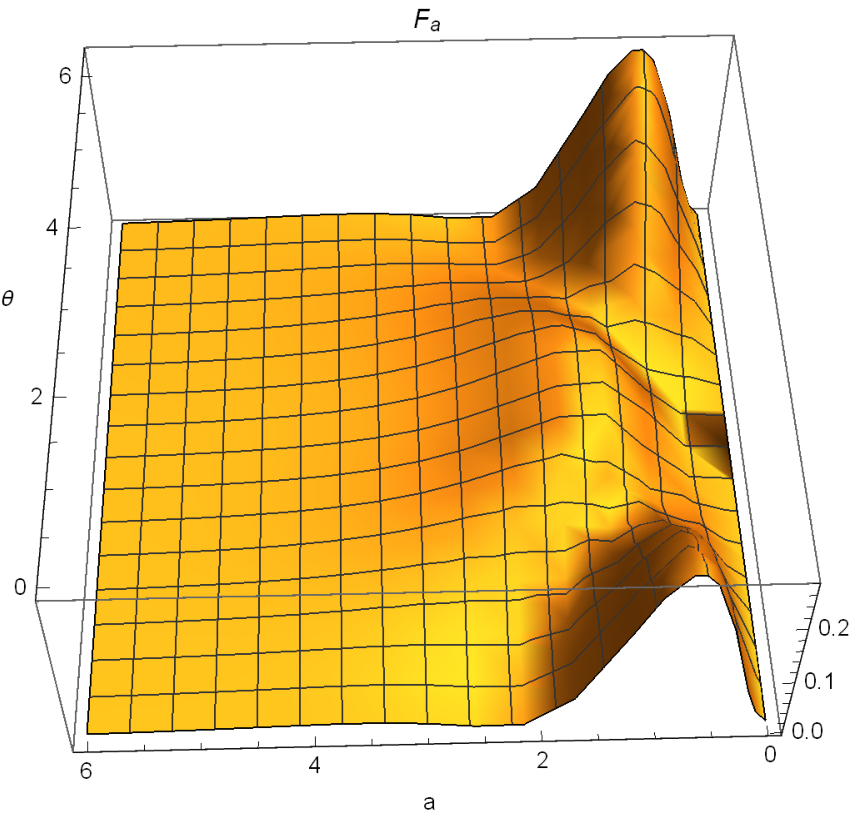}
\includegraphics[scale=0.55]{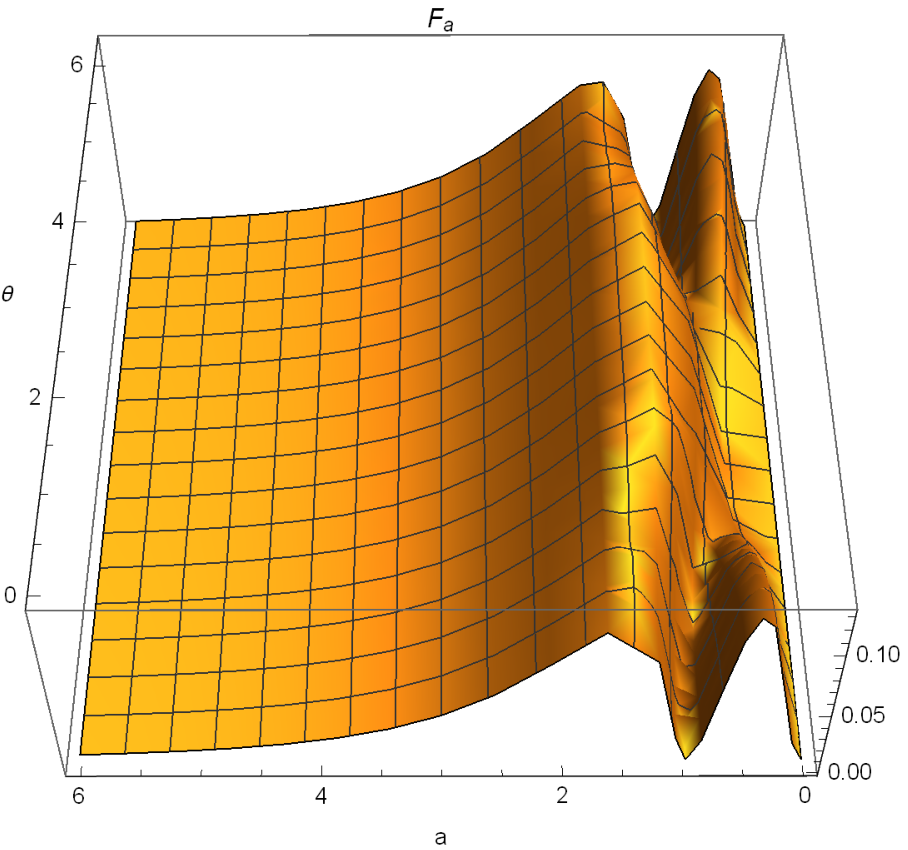}
\includegraphics[scale=0.55]{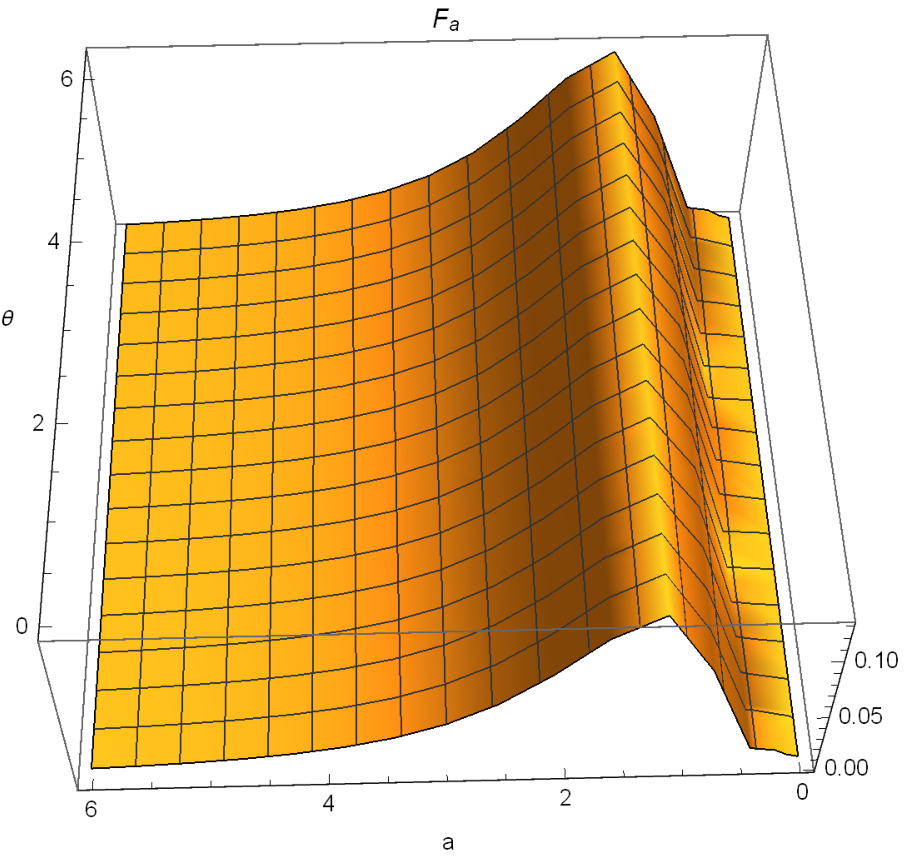}
\caption{\label{fig-Fath-b} The QFI of acceleration as a function of $a$ and $\theta$ for $z=0.5$. We take $\tau=5$ (left panel), $\tau=20$ (middle panel), and $\tau=40$ (right panel). }
\end{centering}
\end{figure}

\begin{figure}[H]
\begin{centering}
\includegraphics[scale=0.55]{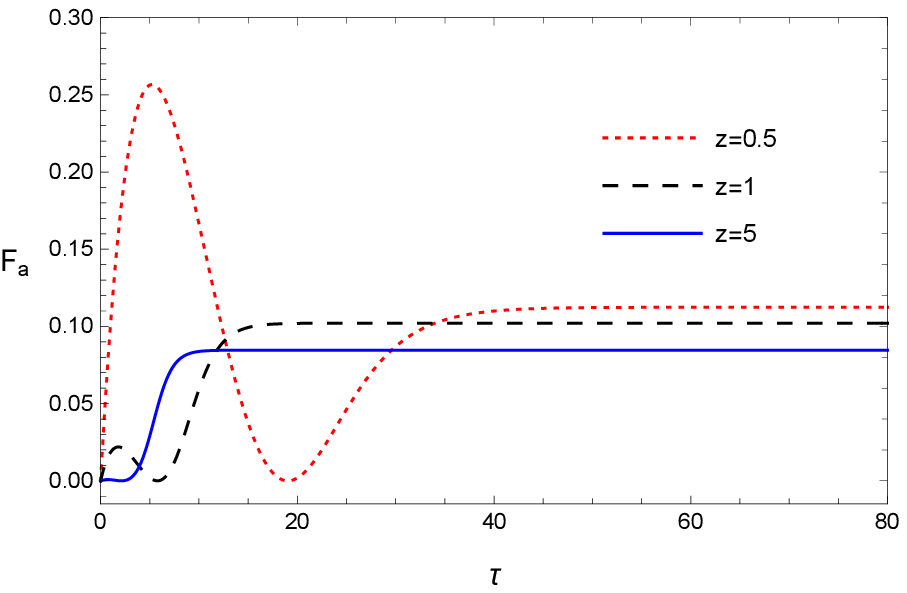}
\includegraphics[scale=0.55]{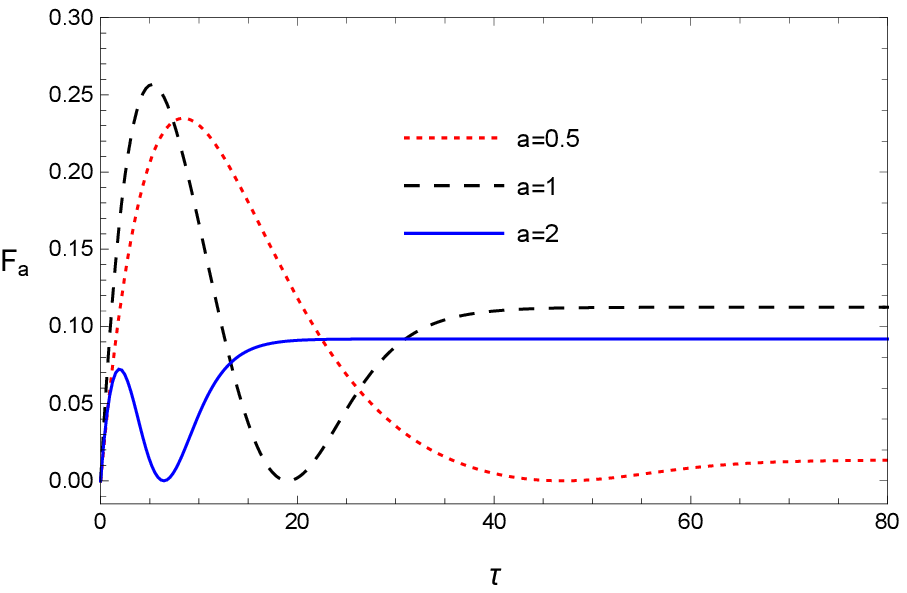}
\includegraphics[scale=0.53]{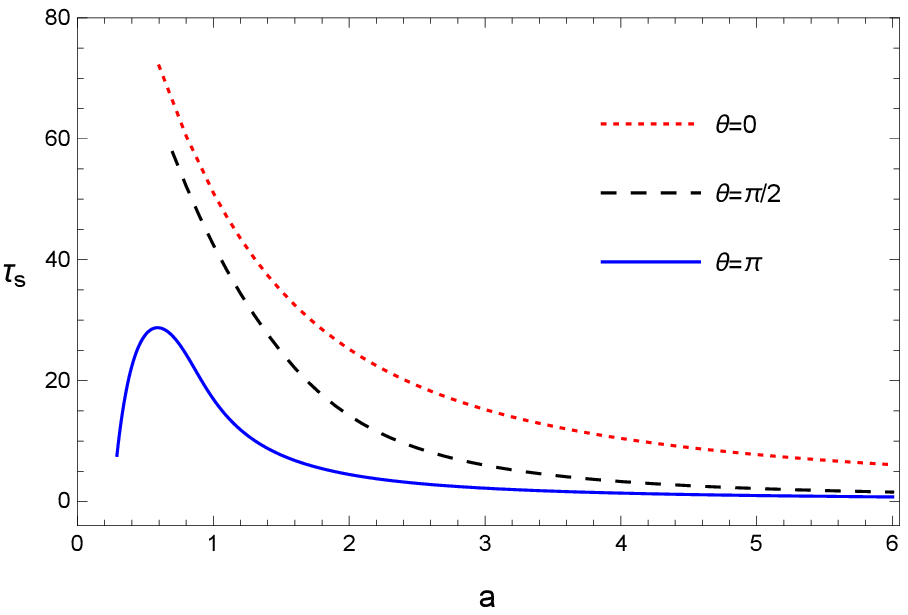}
\includegraphics[scale=0.55]{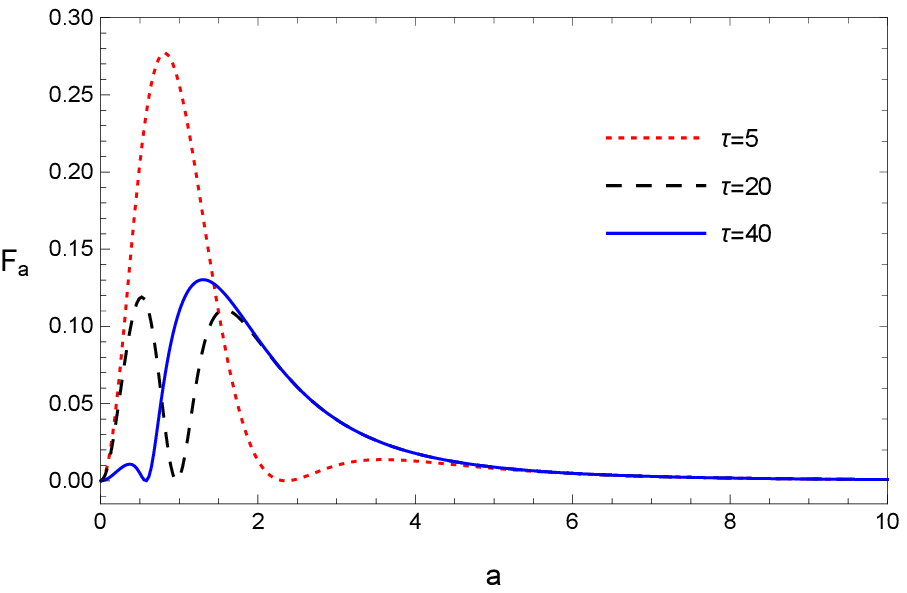}
\includegraphics[scale=0.55]{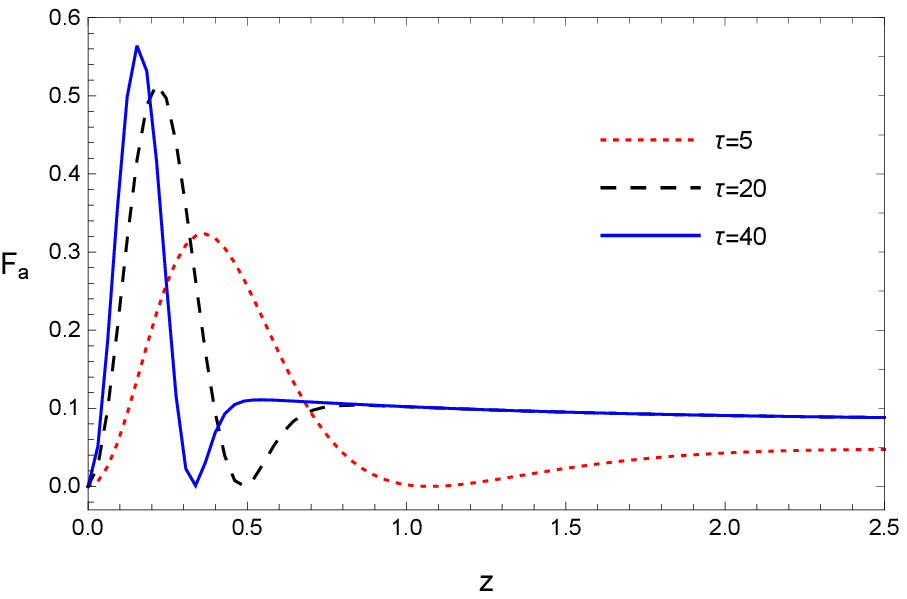}
\includegraphics[scale=0.55]{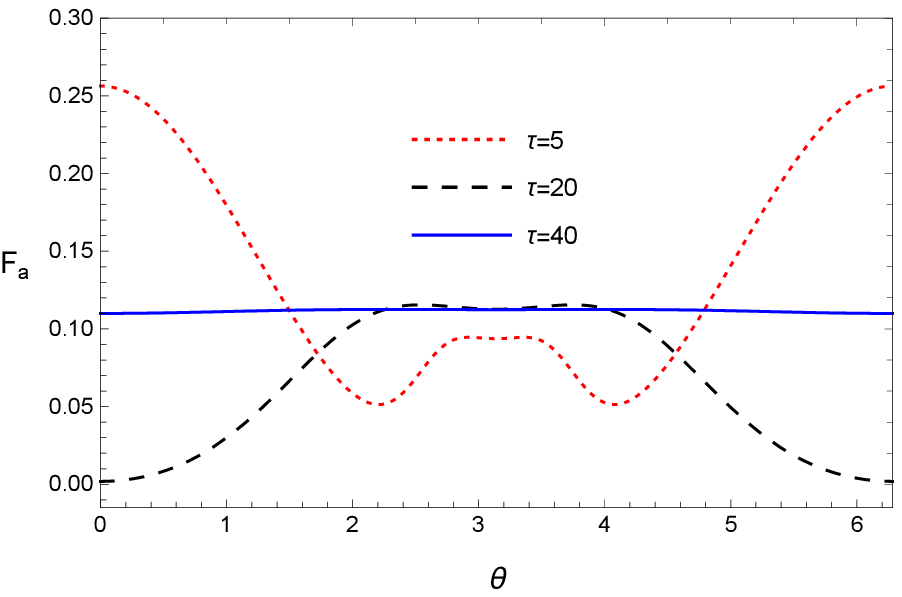}
\caption{\label{b-2dfig-a}The QFI of acceleration as a function of $\tau$ for $\theta=0$ and $a=1$ with different $z$ in the top left panel, as a function of $\tau$ for $\theta=0$ and $z=0.5$ with different $a$ in the top middle panel, as a function of $a$ for $\theta=0$ and $z=0.5$ with different $\tau$ in the bottom left panel, as a function of $z$ for $\theta=0$ and $a=1$ with different $\tau$ in the bottom middle panel, and as a function of $\theta$ for $z=0.5$ and $a=1$ with different $\tau$ in the bottom right panel. The saturation time as a function of $a$ for $z=0.5$ with different $\theta$ by using numerical method in the top right panel.}
\end{centering}
\end{figure}

We describe the QFI of acceleration with $z=0.01$ (left panel), $z=1$ (middle panel), and $z=5$ (right panel) as a function of $a$ and $\tau$ for $\theta=0$ in Fig. \ref{fig-Fat-b}. The QFI fluctuates, and we can obtain the larger peak value of $F_a$ compared to the absence of a boundary. Fixing $\theta=0$, we plot the QFI of acceleration with $\tau=5$, $\tau=100$, and $\tau=80000$ as a function of $a$ and $z$ in Fig. \ref{fig-Faz-b}. The maximum value of $F_a$ is closer to the boundary for a longer time, which also can be seen in the top left and bottom middle panels of Fig. \ref {b-2dfig-a}. In the top left panel of Fig. \ref {b-2dfig-a}, we see that a larger stable value can be obtained at a smaller $z$. In the top middle panel of Fig. \ref {b-2dfig-a}, $F_{a}$ with $a=2$ reaches the stable value faster than $a=1$ and $a=0.5$ for $\theta=0$ and $z=0.5$, but slower than unbounded case. The relation between the saturation time and $a$ is similar to unbound case, just as shown in the top right panel. In the bottom left panel, by the numerical method, these peak values correspond to $a=0.8117$ and $a=3.5863$ for $\tau=5$, $a=0.5207$ and $a=1.5774$ for $\tau=20$, and $a=0.3703$ and $a=1.3034$ for $\tau=40$. For fixed $\tau$, while $a$ is larger than the specific value corresponds the second peak value, the precision reduces. In Fig. \ref{fig-Fath-b}, we depict the QFI of acceleration with $\tau=5$ (left panel), $\tau=20$ (middle panel), and $\tau=40$ (right panel) as a function of $a$ and $\theta$ for fixed $z=0.5$. The QFI displays the periodicity with respect to $\theta$. The variation of QFI with respect to $\theta$ gradually fades away with the evolution of time, but it lasts a longer time than the previous unbounded case. In the bottom right panel of Fig. \ref {b-2dfig-a}, the QFI in the excited state ($\theta=0$ and $\theta=2\pi$) firstly takes the maximum and then takes the minimum, while achieves a stable value after a certain time. The behavior is contrast to that of $F_{a}$ in the unbounded situation. The detection range of the acceleration has been expanded because of adding a boundary.

\section{Quantum estimation of temperature for a static atom immersed in a thermal bath without and with a boundary}

\subsection{Estimation of temperature for a static atom immersed in a thermal bath without boundary}

In this section, we consider a static atom immersed in a thermal bath without boundary, and the field correlation function is given by
\begin{eqnarray}
G^+(t,t')=-\frac{1}{4\pi^2}\Sigma_{m=-\infty}^{\infty}\frac{1}{(t-t'-i m \beta-i \varepsilon)^2}.
\end{eqnarray}

The Fourier transformation of the field correlation function can be expressed as
\begin{eqnarray}\label{fourierT}
{\cal G}^{0}(\lambda)=\frac{1}{2\pi}\frac{\lambda}{1-e^{-\lambda/T}},
\end{eqnarray}
where the temperature is $T={1}/{\beta}$.

The coefficients for the Kossakowski matrix are
\begin{eqnarray}\label{abT}
\begin{aligned}
&A_0=\frac{\mu^2 \omega_0 }{8\pi}\coth\frac{\omega_0}{2T}     \;,\\
&B_0=\frac{\mu^2 \omega_0 }{8\pi} \;.\\
\end{aligned}
\end{eqnarray}
In the following discussion, we use $t\rightarrow \tau\equiv{\mu^2\omega_0}t/{(2\pi)}$ and $T\rightarrow \tilde{T}\equiv{T}/{\omega_0}$. For simplicity, $\tilde{T}$ will be written as $T$. We can obtain the QFI of the temperature $F_T$.

\begin{figure}[H]
\begin{centering}
\includegraphics[scale=0.55]{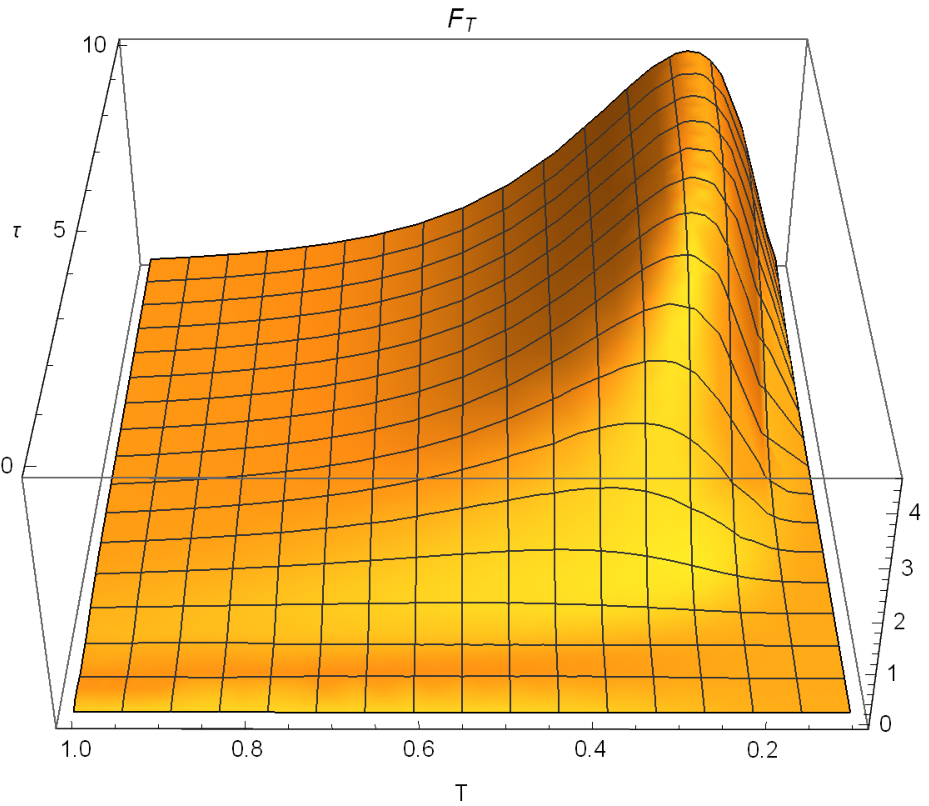}
\includegraphics[scale=0.55]{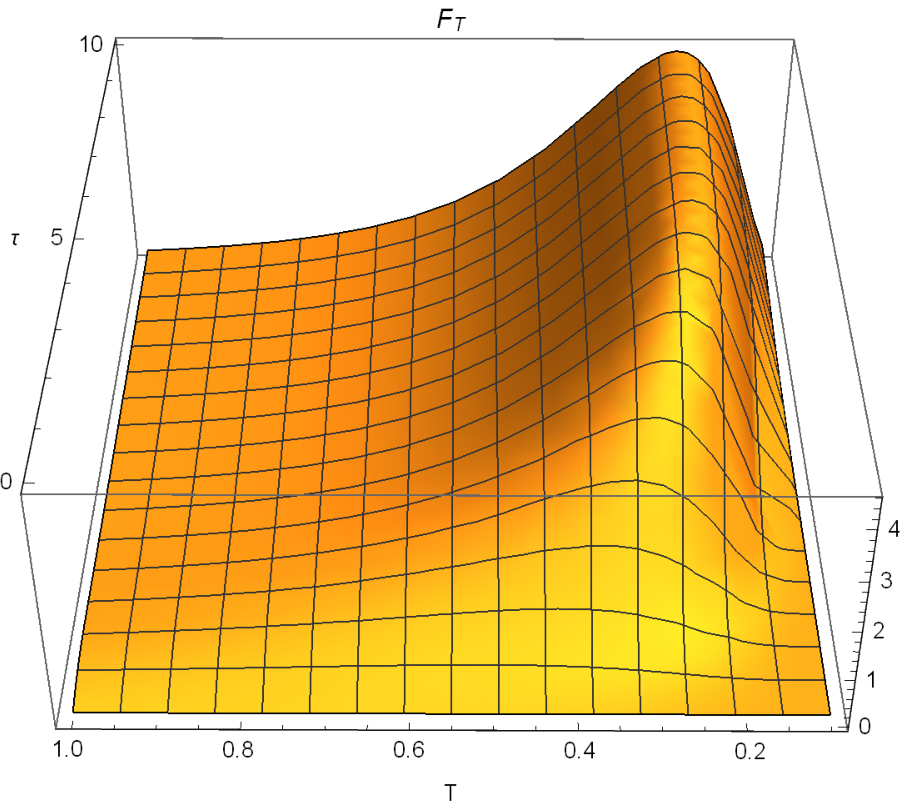}
\includegraphics[scale=0.55]{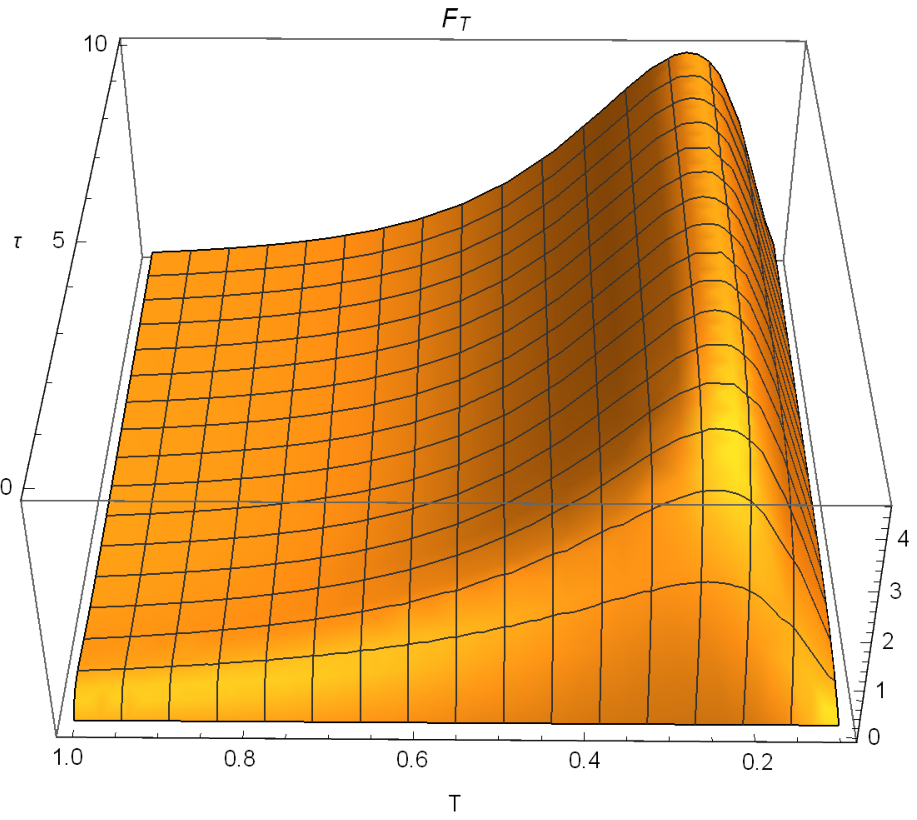}
\caption{\label{fig-FTt} The QFI of temperature as a function of the temperature $T$ and the evolution time $\tau$ for different initial state parameters $\theta$. We take $\theta=0$ (left panel), $\theta=\pi/2$ (middle panel), and $\theta=\pi$ (right panel). }
\end{centering}
\end{figure}

\begin{figure}[H]
\begin{centering}
\includegraphics[scale=0.53]{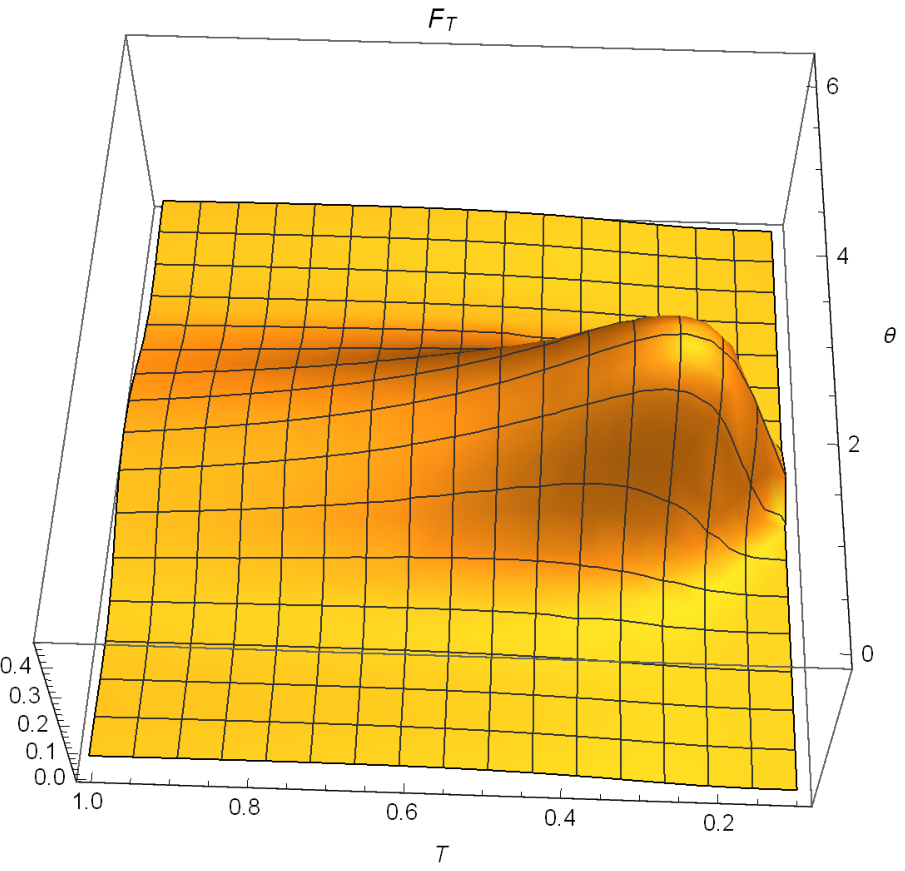}
\includegraphics[scale=0.53]{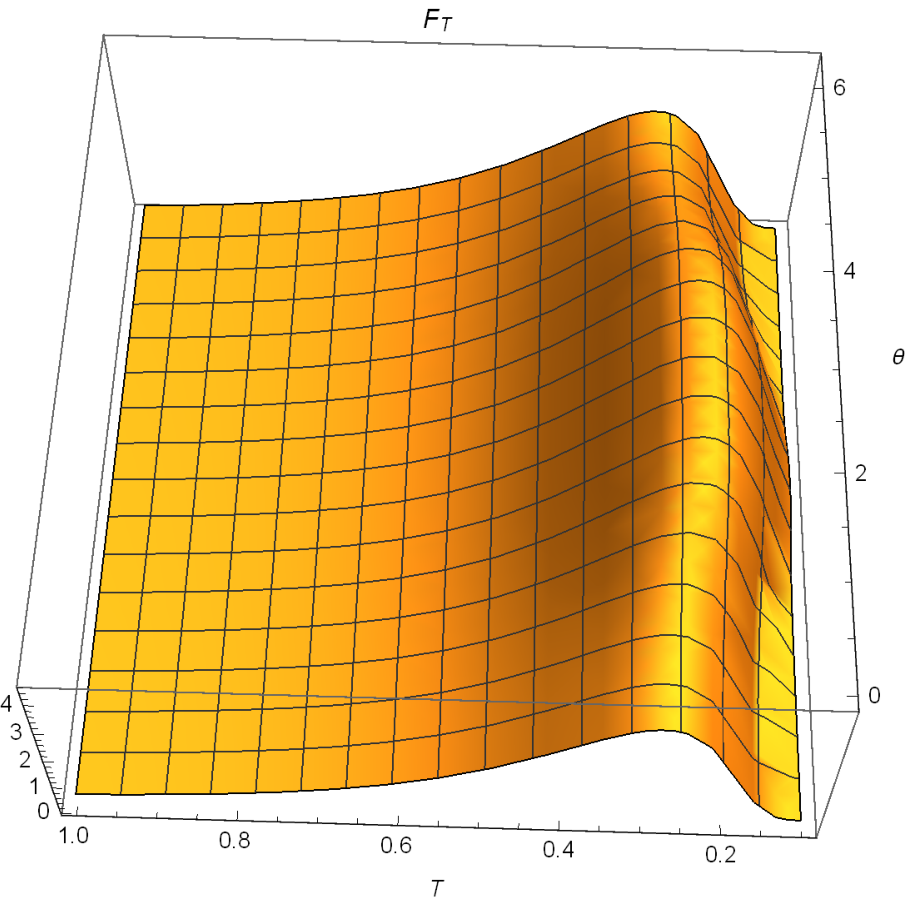}
\includegraphics[scale=0.53]{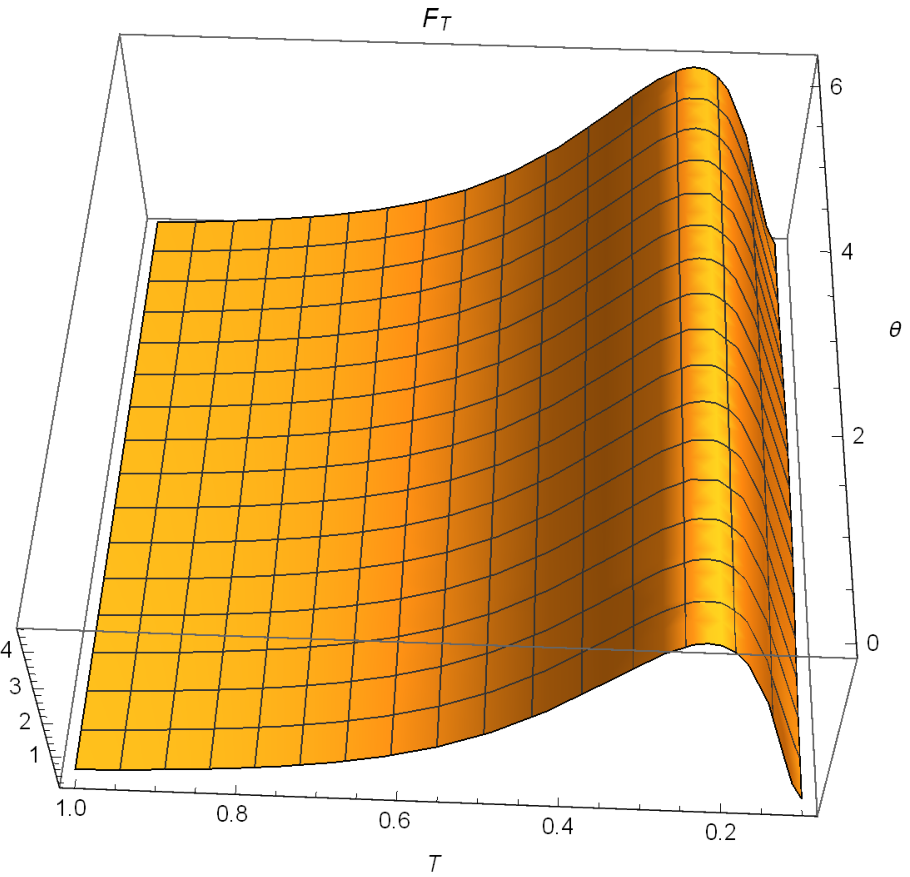}
\caption{\label{fig-FTth} The QFI of temperature as a function of the temperature $T$ and the initial state parameter $\theta$ for different $\tau$. We take $\tau=0.1$ (left panel), $\tau=5$ (middle panel), and $\tau=9$ (right panel). }
\end{centering}
\end{figure}

\begin{figure}[H]
\begin{centering}
\includegraphics[scale=0.75]{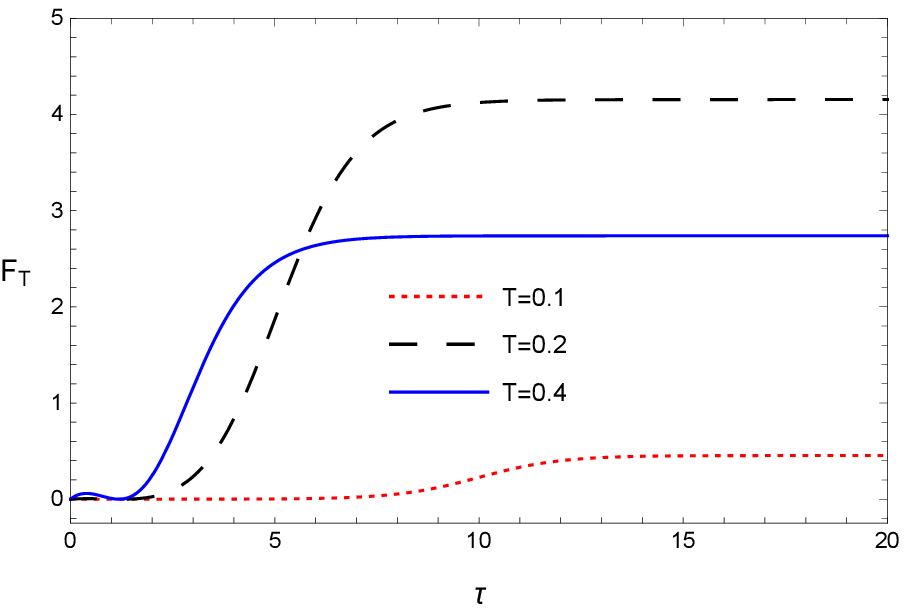}
\includegraphics[scale=0.75]{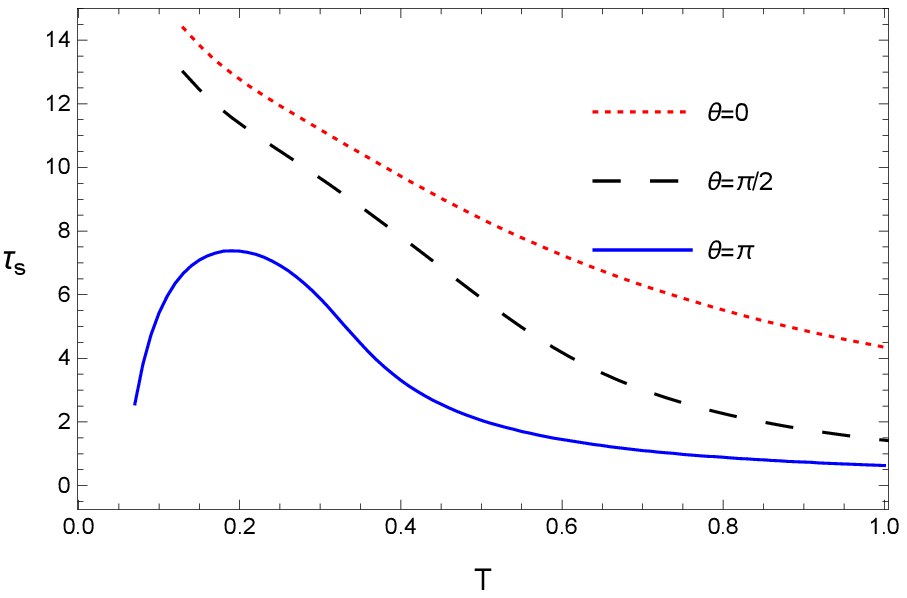}
\includegraphics[scale=0.75]{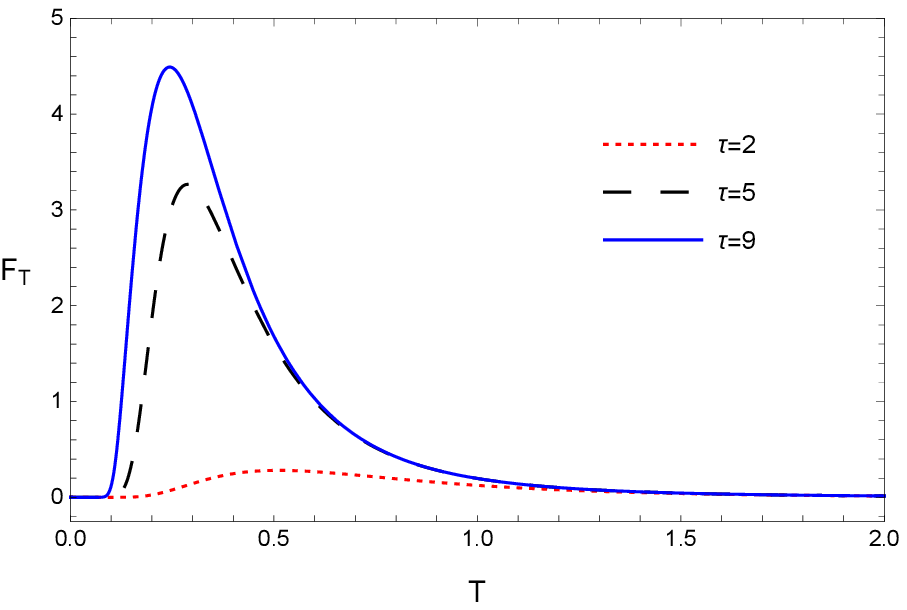}
\includegraphics[scale=0.75]{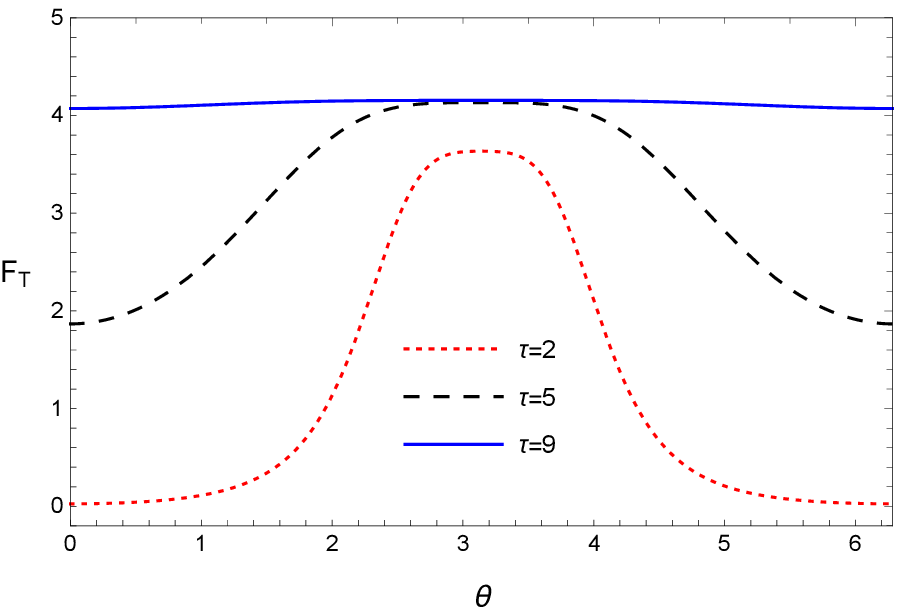}
\caption{\label{2dfig-T}The QFI of temperature as a function of $\tau$ for $\theta=0$ with different $T$ in the top left panel, as a function of $T$ for $\theta=0$ with different $\tau$ in the bottom left panel, and as a function of $\theta$ for $T=0.2$ with different $\tau$ in the bottom right panel. The saturation time as a function of $T$ with different $\theta$ by using numerical method in the top right panel.}
\end{centering}
\end{figure}

We plot the QFI of temperature as a function of $T$ and $\tau$ for $\theta=0$, $\theta=\pi/2$, and $\theta=\pi$ in Fig. \ref {fig-FTt}. There exists a peak value of $F_T$ for a certain time, indicating the optimal precision of estimation can be achieved when choosing the appropriate $T$ and $\tau$. We describe the QFI with different $T$ as a function of $\tau$ for fixed $\theta=0$ in the top left panel of Fig. \ref{2dfig-T}. The QFI with $T=0.4$ reaches the stable value which is also the maximum value quicker than $T=0.1$ and $T=0.2$ for $\theta=0$. As we amplify $T$, the saturation time decreases for $\theta\neq\pi$, but first increases and then decreases for $\theta=\pi$. We present the case of $\theta=0$, $\theta=\pi/2$, and $\theta=\pi$ in the top right panel. In the bottom left panel of Fig. \ref{2dfig-T}, we find that the peak value of QFI increases with the evolution of time, and eventually reaches to the maximum. For fixed $\tau$, while $T$ is larger than the specific value which corresponds to the peak value of QFI, the precision reduces. By the numerical method, we obtain the specific values $T=0.5113$ for $\tau=2$, $T=0.2876$ for $\tau=5$, and $T=0.2439$ for $\tau=9$.  We depict the QFI of temperature as a function of $T$ and $\theta$ for $\tau=0.1$, $\tau=5$, and $\tau=9$ in Fig. \ref {fig-FTth}. From the left panel, we see that different $\theta$ in general leads to the different QFI, and $F_T$ takes peak value at $\theta=\pi$. The difference gradually fades away with the evolution of time as shown in the middle panel and right panel. In the bottom right panel of Fig. \ref{2dfig-T}, beyond a certain time, the QFI will take the maximum for different $\theta$.

\subsection{Estimation of temperature for a static atom immersed in a thermal bath with a boundary}

We consider a static atom immersed in a thermal bath with a boundary, and the field correlation function can be expressed as
\begin{eqnarray}
G^+(t,t')=-\frac{1}{4\pi^2}\Sigma_{m=-\infty}^{\infty}\left[ \frac{1}{(t-t'-i m \beta-i \varepsilon)^2}- \frac{1}{(t-t'-i m \beta-i \varepsilon)^2-(2z)^2}\right].
\end{eqnarray}

The Fourier transformation of the field correlation function is
\begin{eqnarray}\label{fourierT}
{\cal G}(\lambda)=\frac{1}{2\pi}\frac{\lambda}{1-e^{-\lambda/T}}-\frac{1}{2\pi}\frac{\lambda}{1-e^{-\lambda/T}}\frac{\sin(2z \lambda)}{2z\lambda},
\end{eqnarray}
with the temperature $T={1}/{\beta}$.

The coefficients for the Kossakowski matrix are given by
\begin{eqnarray}\label{abT}
\begin{aligned}
&A_{b}=\frac{\mu^2 \omega_0 }{8\pi}\coth\frac{\omega_0}{2T}\left[1-\frac{\sin(2z\omega_0)}{2z\omega_0}\right]     \;,\\
&B_{b}=\frac{\mu^2 \omega_0 }{8\pi}\left[1-\frac{\sin(2z\omega_0)}{2z\omega_0}\right] \;.\\
\end{aligned}
\end{eqnarray}
In the following discussion, we use $t\rightarrow \tau\equiv{\mu^2\omega_0}t/{(2\pi)}$, $T\rightarrow \tilde{T}\equiv{T}/{\omega_0}$, and $z\rightarrow \tilde{z}\equiv z \omega_0$. For simplicity, $\tilde{T}$ and $\tilde{z}$ will be written as $T$ and $z$. We can obtain the QFI of the temperature $F_T$.

\begin{figure}[H]
\begin{centering}
\includegraphics[scale=0.52]{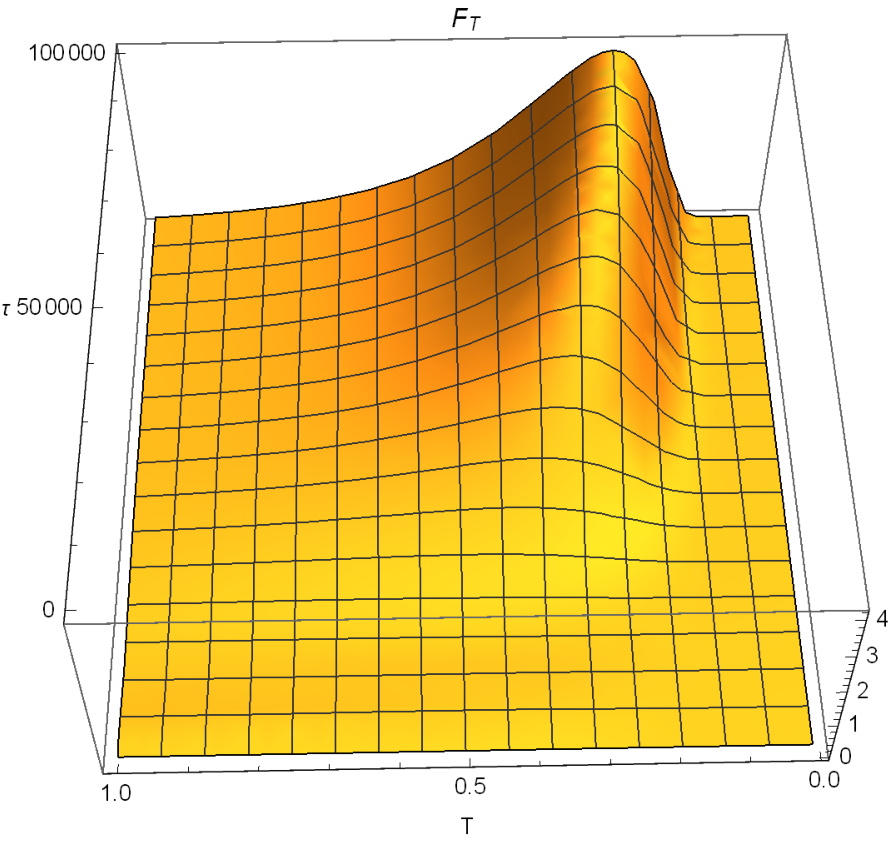}
\includegraphics[scale=0.52]{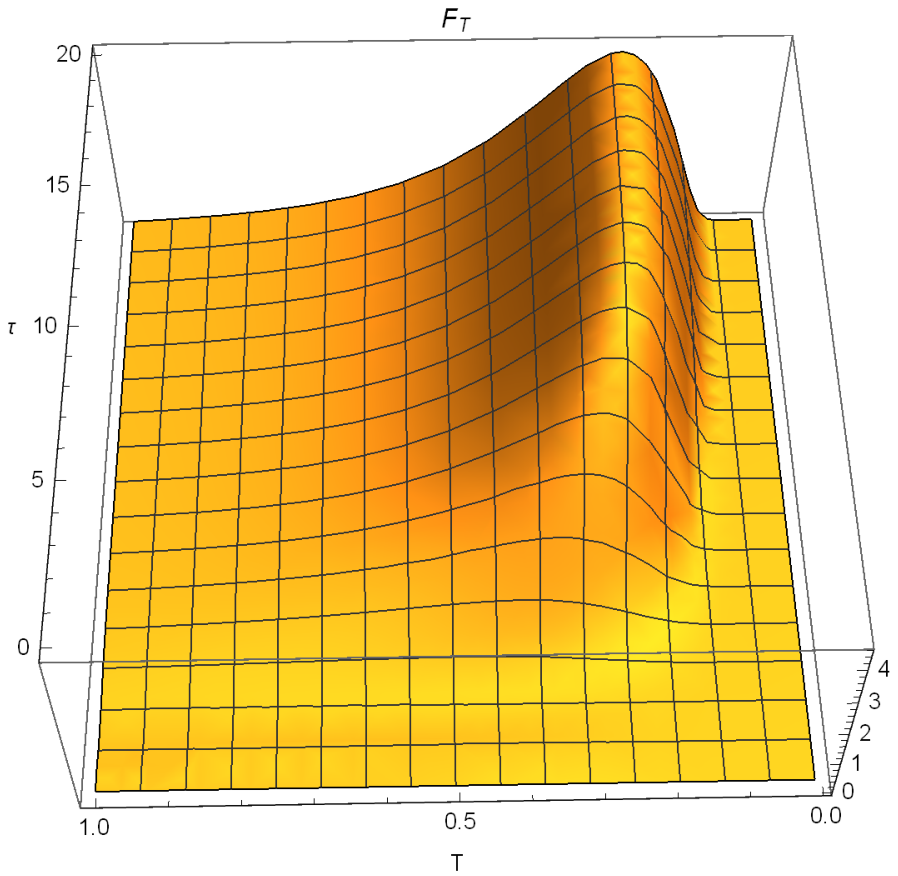}
\includegraphics[scale=0.53]{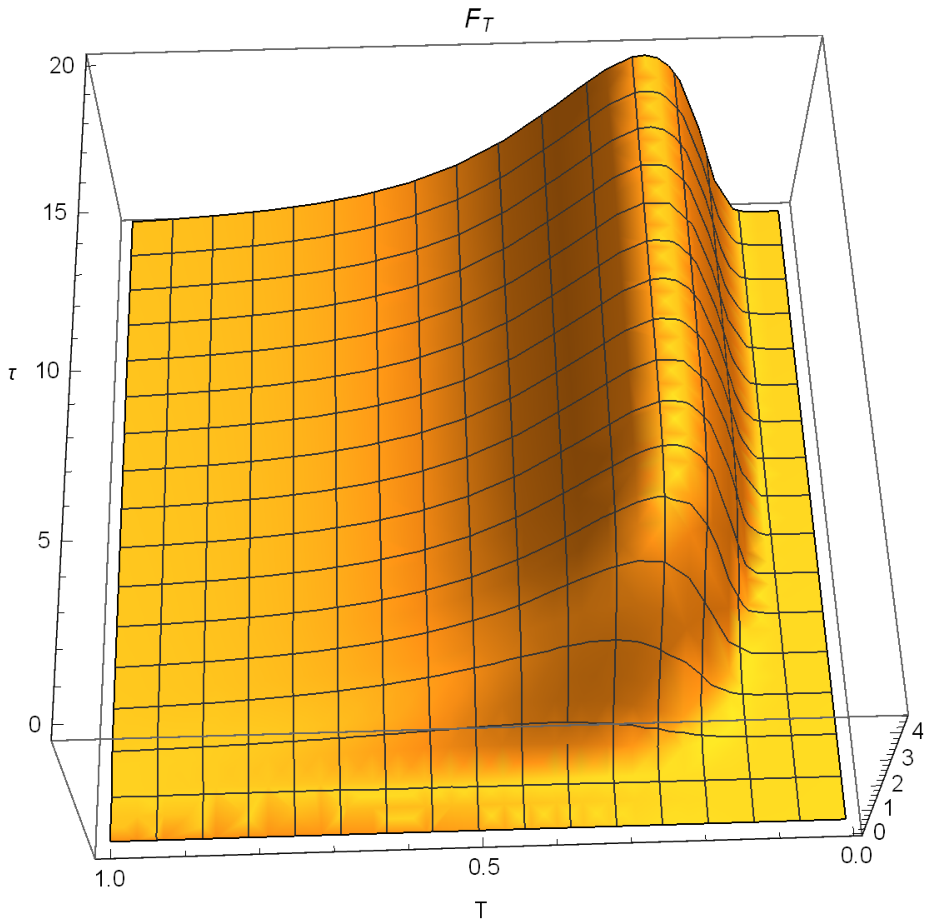}
\caption{\label{fig-FTt-b} The QFI of temperature as a function of $T$ and $\tau$ for $\theta=0$. We take $z=0.01$ (left panel), $z=1$ (middle panel), and $z=5$ (right panel). }
\end{centering}
\end{figure}

\begin{figure}[H]
\begin{centering}
\includegraphics[scale=0.55]{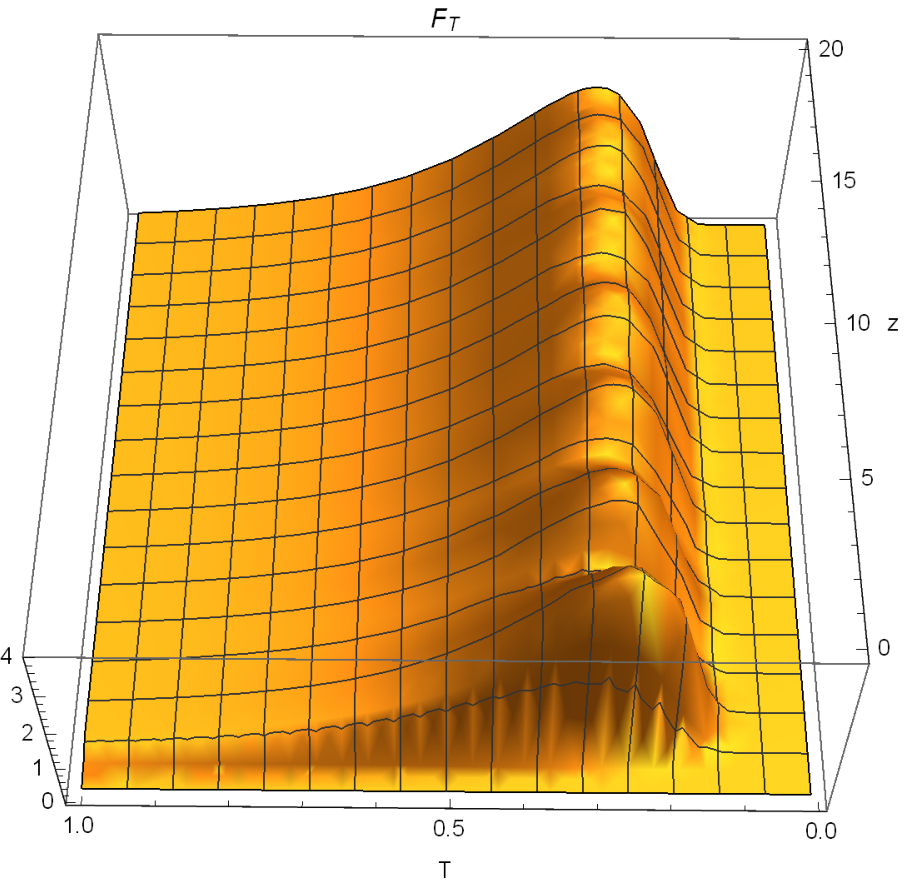}
\includegraphics[scale=0.55]{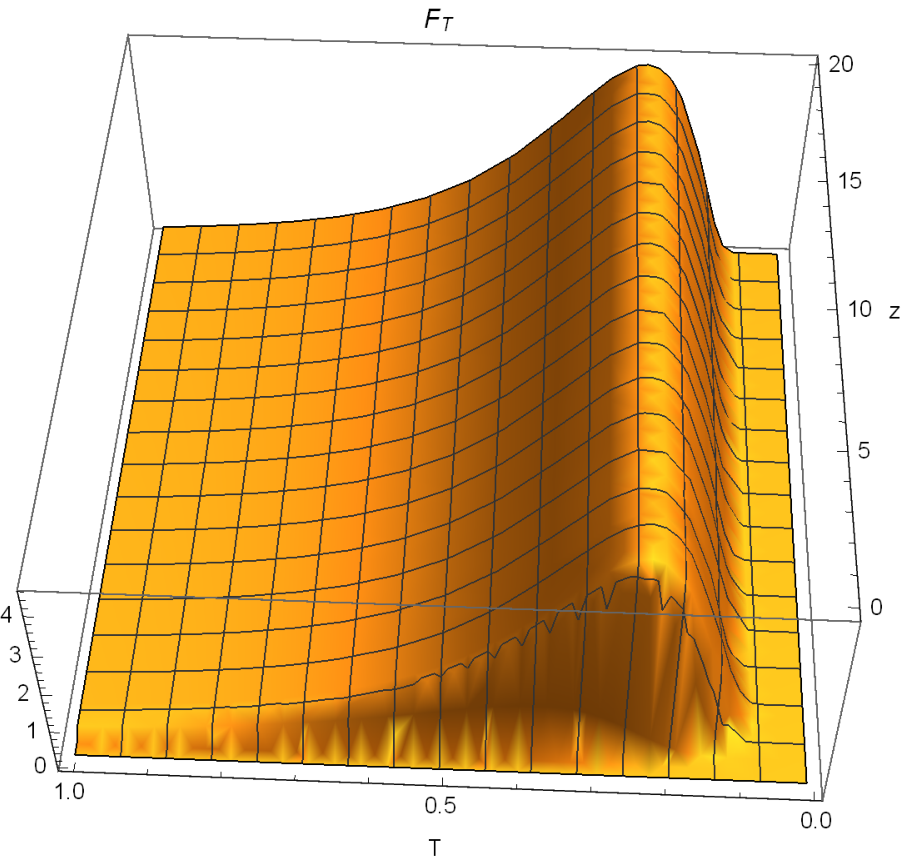}
\includegraphics[scale=0.55]{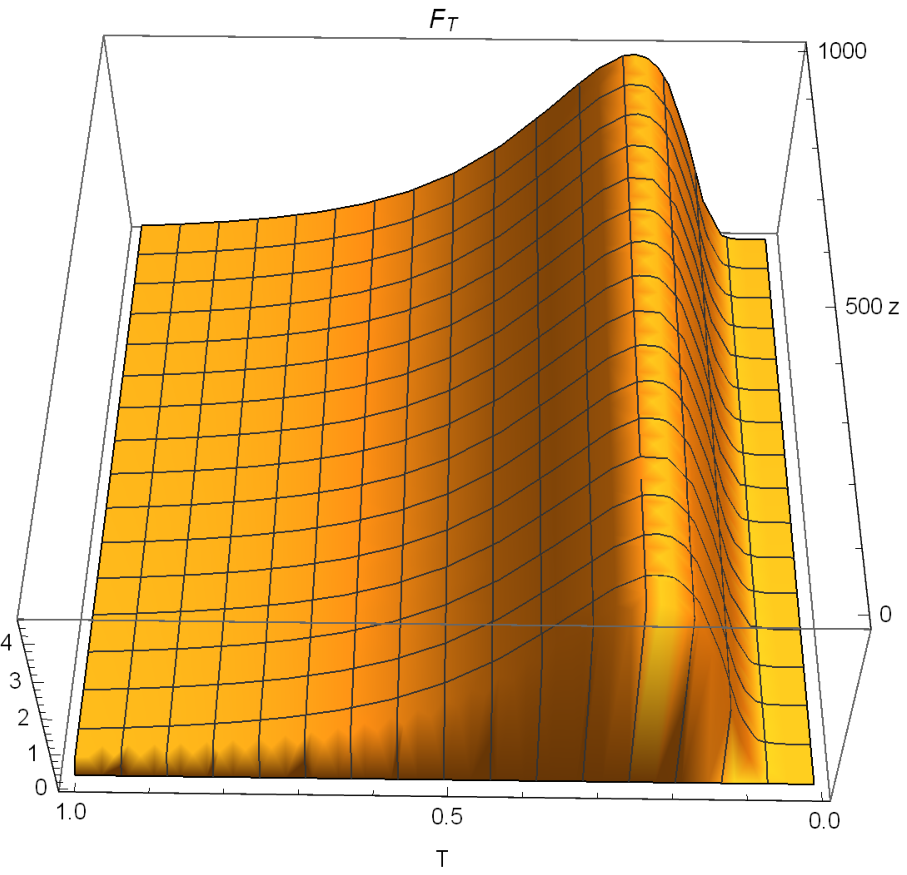}
\caption{\label{fig-FTz-b} The QFI of temperature as a function of $T$ and $z$ for $\theta=0$. We take $\tau=5$ (left panel), $\tau=10$ (middle panel), and $\tau=80000$ (right panel). }
\end{centering}
\end{figure}

\begin{figure}[H]
\begin{centering}
\includegraphics[scale=0.53]{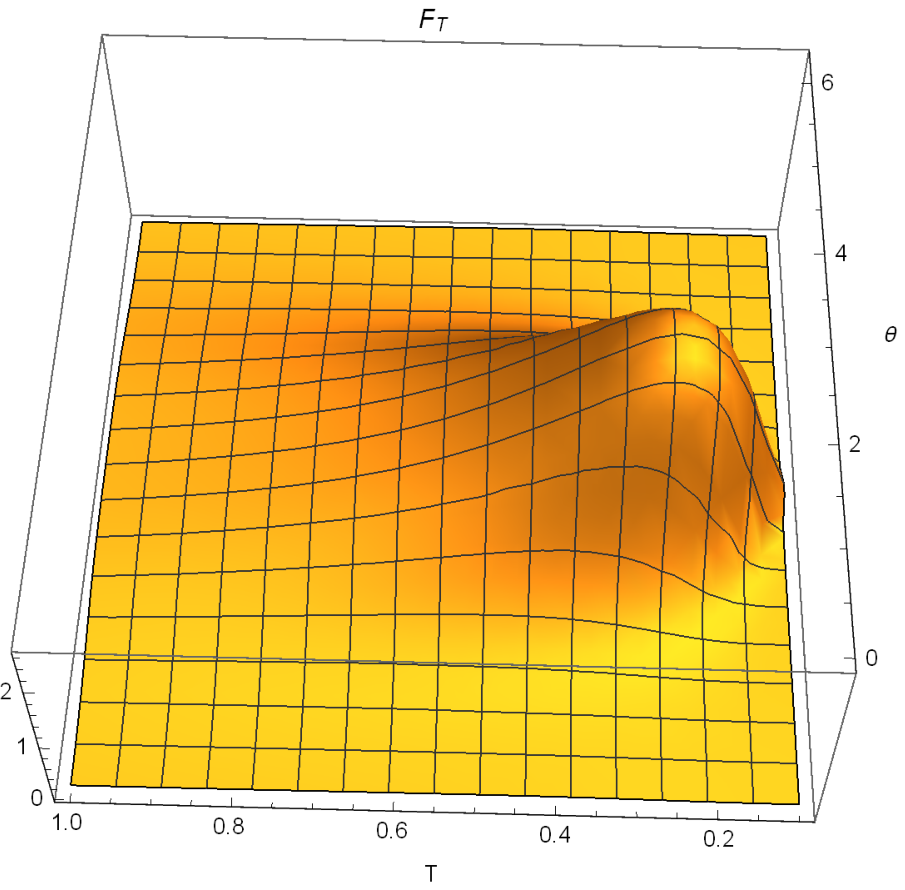}
\includegraphics[scale=0.53]{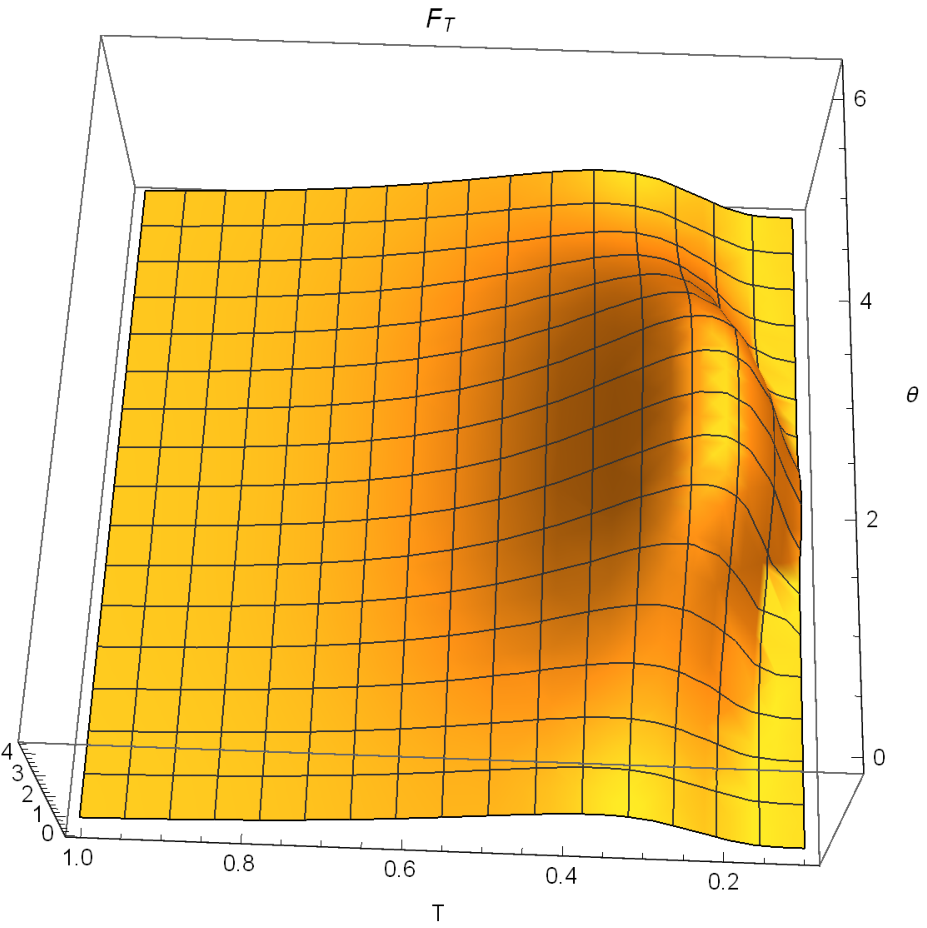}
\includegraphics[scale=0.65]{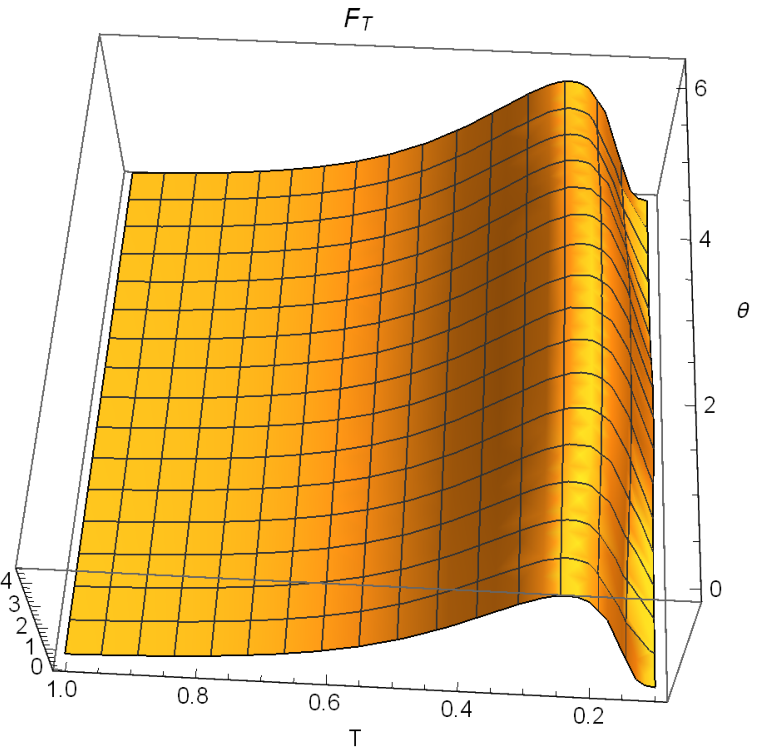}
\caption{\label{fig-FTth-b} The QFI of temperature as a function of $T$ and $\theta$ for $z=0.5$. We take $\tau=5$ (left panel), $\tau=20$ (middle panel), and $\tau=40$ (right panel). }
\end{centering}
\end{figure}

\begin{figure}[H]
\begin{centering}
\includegraphics[scale=0.55]{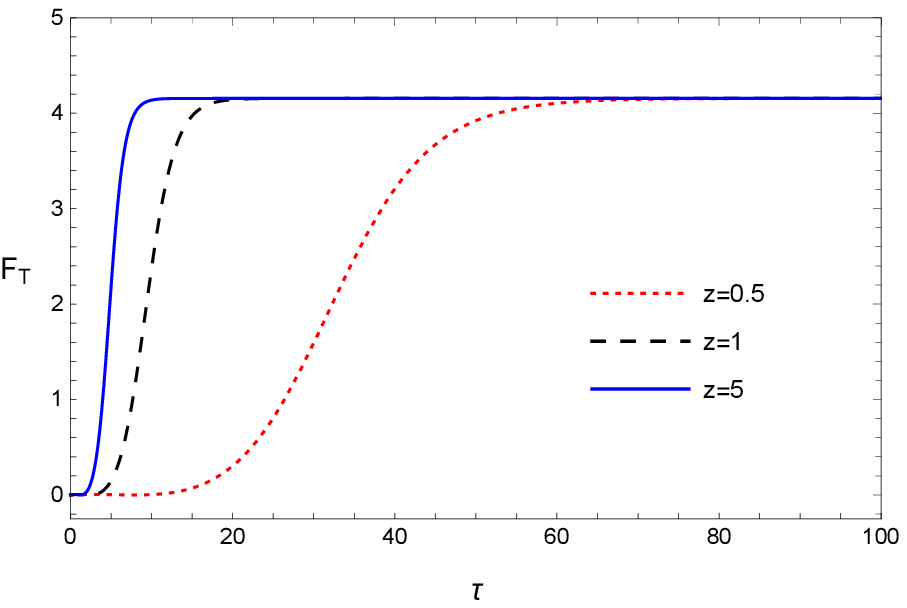}
\includegraphics[scale=0.55]{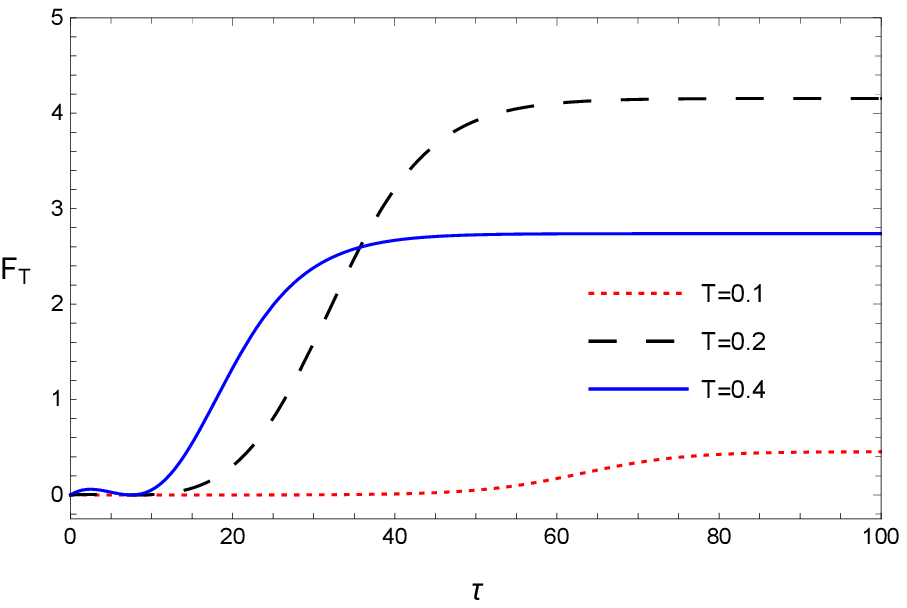}
\includegraphics[scale=0.55]{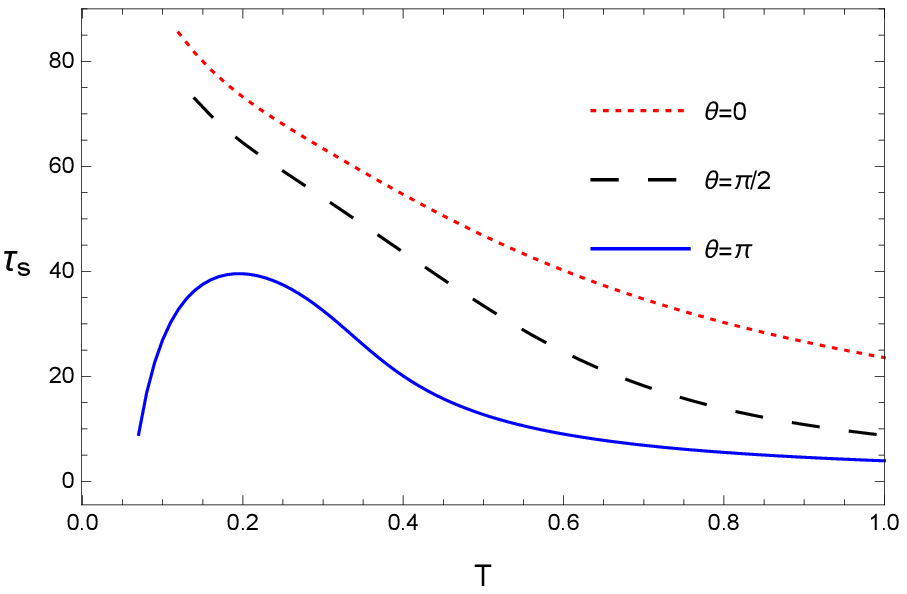}
\includegraphics[scale=0.55]{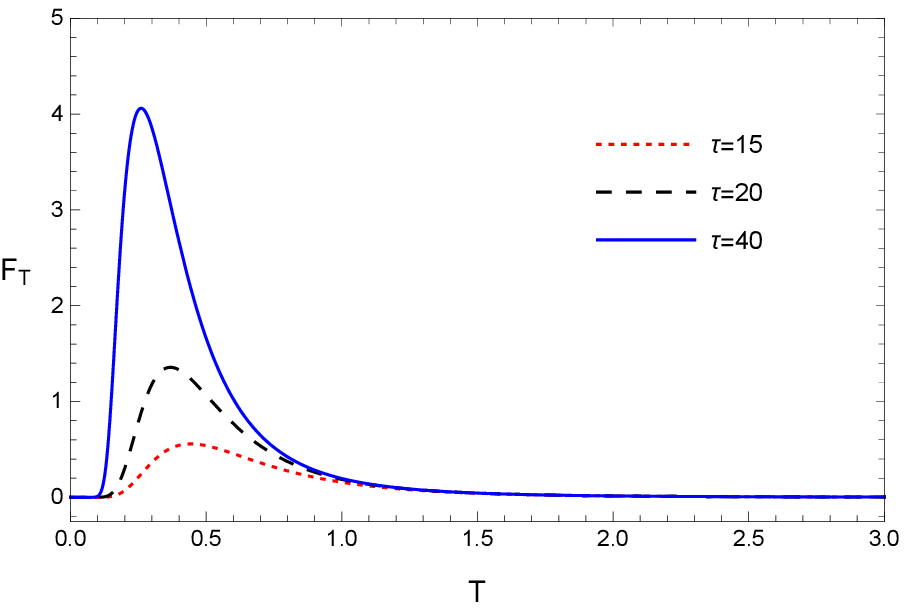}
\includegraphics[scale=0.55]{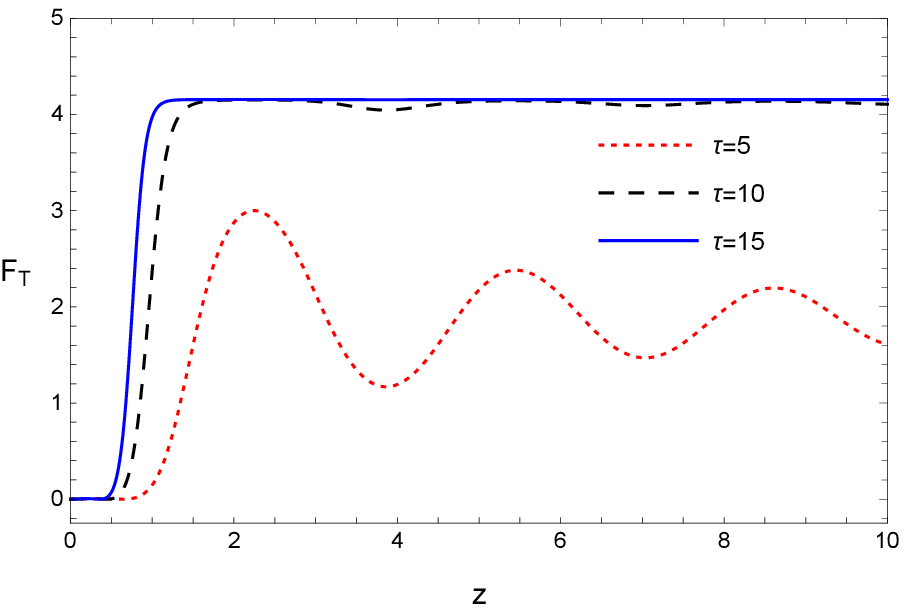}
\includegraphics[scale=0.55]{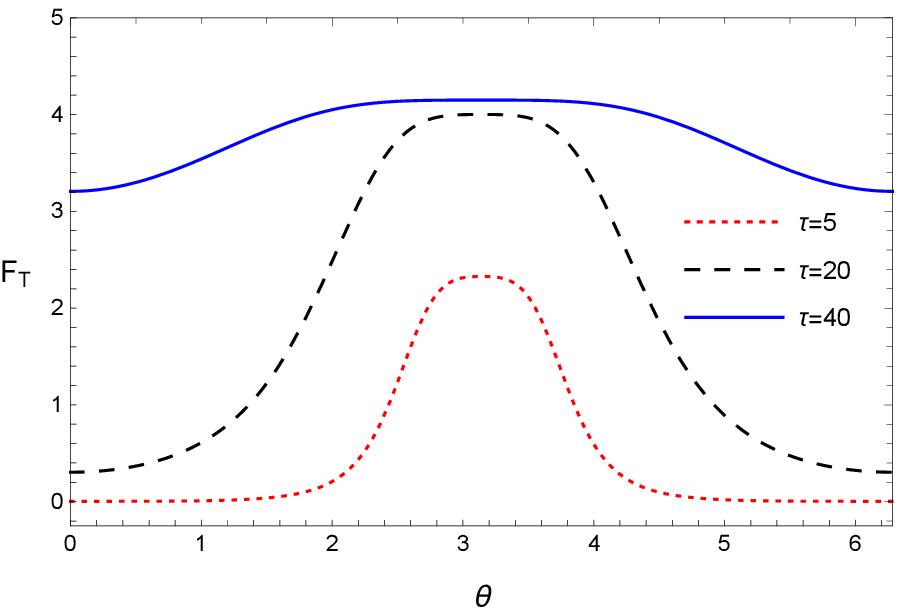}
\caption{\label{b-2dfig-T}The QFI of temperature as a function of $\tau$ for $\theta=0$ and $T=0.2$ with different $z$ in the top left panel, as a function of $\tau$ for $\theta=0$ and $z=0.5$ with different $T$ in the top middle panel, as a function of $T$ for $\theta=0$ and $z=0.5$ with different $\tau$ in the bottom left panel, as a function of $z$ for $\theta=0$ and $T=0.2$ with different $\tau$ in the bottom middle panel, and as a function of $\theta$ for $z=0.5$ and $T=0.2$ with different $\tau$ in the bottom right. The saturation time as a function of $T$ for $z=0.5$ with different $\theta$ by using numerical method in the top right panel.}
\end{centering}
\end{figure}

We present the QFI of temperature with $z=0.01$ (left panel), $z=1$ (middle panel), and $z=5$ (right panel) as a function of $T$ and $\tau$ for fixed $\theta=0$ in Fig. \ref{fig-FTt-b}. Added a boundary, $F_T$ takes a longer time to reach the maximum with the smaller $z$. We plot the QFI of temperature with $\tau=5$ (left panel), $\tau=10$ (middle panel), and $\tau=80000$ (right panel) as a function of $T$ and $z$ for $\theta=0$ in Fig. \ref {fig-FTz-b}. In a very short time, we see that $F_{T}$ fluctuates along $z$ from the left panel, and the fluctuation fades rapidly as shown in the middle panel. The maximum value of $F_T$ can be obtained far away from the boundary for a longer time. In the top left panel of Fig. \ref {b-2dfig-T}, the maximum of QFI is same towards different $z$. The QFI with $T=0.4$ achieves the stable value which is also the maximum value faster than $T=0.1$ and $T=0.2$ for $\theta=0$ and $z=0.5$ in the top middle panel of Fig. \ref {b-2dfig-T}, but slower than unbounded circumstance. The relation between the saturation time and $T$ is similar to unbounded case, just as shown in the top right panel. In the bottom left panel of Fig. \ref {b-2dfig-T}, for fixed $\tau$, while $T$ is larger than the specific value which corresponds to the peak value of QFI, the precision reduces. By the numerical method, we obtain the specific values $T=0.4432$ for $\tau=15$, $T=0.3683$ for $\tau=20$, and $T=0.2604$ for $\tau=40$. In the bottom middle panel of Fig. \ref {b-2dfig-T}, the QFI has a large fluctuation with respect to $z$ in a short time. After a certain time, the QFI reaches the maximum quickly. We depict the QFI of temperature with $\tau=5$ (left panel), $\tau=20$ (middle panel), and $\tau=40$ (right panel) as a function of $T$ and $\theta$ for $z=0.5$ in Fig. \ref {fig-FTth-b}. In a very short time, the QFI varies with the initial state parameter $\theta$. The variation of QFI with respect to $\theta$ lasts a longer time than unbounded case. From the bottom right panel of Fig. \ref {b-2dfig-T}, $F_T$ firstly takes peak value at $\theta=\pi$, and will take the maximum with different $\theta$ after a certain time, which is in strong contrast to the behavior of $F_a$ with a boundary.

\section{conclusion}

In the open quantum systems, we have investigated the QFI of acceleration for a circularly accelerated two-level atom coupled to a scalar field in the Minkowski vacuum without and with a reflecting boundary in the ultra-relativistic limit. With increase of $a$, the saturation time decreases for $\theta\neq\pi$, but first increases and then decreases for $\theta=\pi$. Without a boundary, there exists a peak value of QFI with a certain time, which indicates the optimal precision of estimation can be achieved when choosing an appropriate range. $F_a$ is the periodic function of the initial state parameter $\theta$, and firstly takes peak value in the ground state of the atom. However, the variation of QFI with respect to $\theta$ gradually fades away with the evolution of time. With a boundary, we can obtain a larger peak value of QFI compared to the unbounded case. The detection range of acceleration has been expanded, which indicates the QFI is protected by the boundary. The maximum value of $F_a$ is closer to the boundary for a longer time. The QFI in the excited state of the atom firstly takes the maximum and then takes the minimum in this case, while it reaches a stable value for different initial states after a certain time. The behavior is in strong contrast to that in the unbounded case.

Besides, we have studied the QFI of temperature for a static atom immersed in a thermal bath without and with a boundary. The relation between the saturation time and $T$ is similar to $a$. Without a boundary, the behavior of $F_T$ shows the similarities with $F_a$. With a boundary, $F_T$ takes a longer time to reach the maximum with the smaller $z$. In a short period, we found that $F_{T}$ fluctuates along $z$, but the fluctuation vanishes quickly. The maximum for the QFI of temperature is same towards different $z$. $F_T$ firstly takes peak value in the ground state of the atom, which is different from the behavior of $F_a$ with a boundary. The results provide references for the relevant experiments.

\begin{acknowledgments}

This work was supported by the National Natural Science Foundation of China under Grants No. 11705144, No. 11775076, and No. 11875025.

\end{acknowledgments}

\end{document}